\documentclass[superscriptaddress,twocolumn]{aastex631}
\usepackage{CJK}
\usepackage{url}
\usepackage{amsmath}
\usepackage[super]{nth}
\usepackage{microtype}
\usepackage{graphicx}

\begin{document}
\begin{CJK*}{UTF8}{gbsn}

\title{Atmospheric Waves Driving Variability and Cloud Modulation on a Planetary-Mass Object}

\author[0000-0002-4831-0329]{Michael K. Plummer}
\affiliation{Department of Physics and Meteorology, United States Air Force Academy, 2354 Fairchild Drive, CO 80840}
\affiliation{Department of Astronomy, The Ohio State University, 140 W. 18th Ave., Columbus, OH 43210, USA}

\correspondingauthor{Michael K. Plummer}
\email{michaelkplummer@gmail.com}

\author[0000-0002-4361-8885]{Ji Wang (王吉)}
\affiliation{Department of Astronomy, The Ohio State University, 140 W. 18th Ave., Columbus, OH 43210, USA} 

\author[0000-0003-3506-5667]{\'Etienne Artigau}
\affiliation{D\'epartment de Physique, Universit\'e de Montr\'eal, IREX, Montr\'eal, QC, H3C 3J7, Canada}
\affiliation{Observatoire du Mont-M\'egantic, Universit\'e de Montr\'eal, Montr\'eal, QC, H3C 3J7, Canada}

\author{Ren\'e Doyon}
\affiliation{Institut Trottier de recherche sur les exoplan\`etes, D\'epartment de Physique, Universit\'e de Montr\'eal, Canada}
\affiliation{Observatoire du Mont-M\'egantic, Universit\'e de Montr\'eal, Montr\'eal, QC, H3C 3J7, Canada}

\author[0000-0002-2011-4924]{Genaro Su\'arez}
\affiliation{Department of Astrophysics, American Museum of Natural History, Central Park West at 79th Street, NY 10024, USA}

% \shorttitle{Atmospheric Waves Driving Variability and Cloud Modulation on a Planetary-Mass Object}
\shorttitle{Modeling SIMP\,J0136+0933 Variability}
\shortauthors{Plummer et al.}

\begin{abstract}

% Due to similar temperatures and spectral characteristics, planetary-mass objects (PMOs) and brown dwarfs (BDs) have long been seen as high signal-to-noise analogs for the study of gas giant exoplanets and their atmospheres. 

Planetary-mass objects and brown dwarfs at the transition ($\rm{T}_{eff}\sim1300$\,K) from relatively red L dwarfs to bluer mid-T dwarfs show enhanced spectrophotometric variability. Multi-epoch observations support atmospheric planetary-scale (Kelvin or Rossby) waves as the primary source of this variability; however, large spots associated with the precipitation of silicate and metal clouds have also been theorized and suggested by Doppler imaging. We applied both wave and spotted models to fit near-infrared (NIR), multi-band ($Y$/$J$/$H$/$K$) photometry of SIMP\,J013656.5+093347 (hereafter SIMP0136), collected at the Canada-France-Hawaii Telescope using the Wide-field InfraRed Camera. SIMP0136 is a planetary-mass object (12.7$\pm1.0 \ \rm{M_J}$) at the L/T transition (T2$\pm0.5$) known to exhibit light curve evolution over multiple rotational periods. We measure the maximum peak-to-peak variability of $6.17\pm0.46\%$, $6.45\pm0.33\%$, $6.51\pm0.42\%$, and $4.33\pm0.38\%$ in the $Y$, $J$, $H$, and $K$ bands respectively, and find evidence that wave models are preferred for all four NIR bands. Furthermore, we determine the spot size necessary to reproduce the observed variations is larger than the Rossby deformation radius and Rhines scale, which is unphysical. Through the correlation between light curves produced by the waves and associated color variability, we find evidence of planetary-scale, wave-induced cloud modulation and breakup, similar to Jupiter's atmosphere and supported by general circulation models. We also detect a $93.8^{\circ}\pm7.4^{\circ}$ ($12.7\sigma$) phase shift between the $H-K$ and $J-H$ color time series, providing evidence for complex vertical cloud structure in SIMP0136's atmosphere. 

\end{abstract}


\section{Introduction}

Brown dwarfs and planetary-mass objects, either isolated or orbiting host stars at wide separations, have complex atmospheric dynamics and chemistry. Due to their similar temperatures, masses, and chemical compositions \citep{Burrows2001}, they serve as analogs to lower mass or more closely orbiting gas giant exoplanets. 

Spectrophotometric variability suggests active atmospheric dynamics in substellar objects. Silicate clouds have been found to form in early L dwarfs and thicken throughout the mid-L spectral class \citep{Suarez&Metchev2022}. Brown dwarfs at effective temperatures ($\rm{T_{eff}}$) of ${\sim} 1300$\,K ~\citep[][and references therein]{Kirkpatrick2005} transition from the mineral cloud-rich, late-L dwarfs \citep{Tsuji1996b,Burrows2006} to the relatively cloud-free, mid-T dwarfs \citep{Burrows&Sharp1999,Tsuji&Nakajima2003,Knapp2004,Cushing2006}. These L/T transition dwarfs exhibit elevated variability \citep{Radigan2014a,Radigan2014b,Eriksson2019,Liu2024}. Younger brown dwarfs with lower surface gravity  and planetary-mass objects also show higher variability rates than field brown dwarfs \citep{Vos2022,Liu2024}. Enhanced L/T transition variability has been historically associated with the precipitation and breakup of mineral cloud decks formed during the L class as the brown dwarf cools towards T class temperatures ~\citep[e.g.,][]{ackerman&marley01,Burgasser2002,Reiners&Basri2008}.  

% Because brown dwarfs, by definition, are below the minimum mass for hydrogen fusion (${\sim}80\,M_J$), they gradually cool throughout their lifetimes. From ${\sim} \rm{}2500\,K \ to \ 1300\,K$ (L dwarfs), their NIR colors grow increasingly red. This is theorized to be due to the formation of mineral clouds (e.g. silicates, iron) in their atmospheres. At ${\sim}1300$\,K measured brown dwarf color shift dramatically blueward such that by ${\sim}700$\,K, their atmospheres are considered to be relatively cloud-free. In between is the L/T transition where it is thought that the mineral clouds formed during the L class precipitate out of brown dwarf atmospheres.

% As brown dwarfs transition from late L dwarfs, with silicate and metal clouds reddening near-infrared color indices, to mid-T dwarfs, with bluer colors and relatively cloud-free skies, enhanced spectrophotometric variability has been observed \citep{}

Two primary atmospheric dynamical structures are fueling the enhanced variability according to general circulation models (GCMs): planetary-scale waves (e.g., Kelvin and Rossby waves) and vortices  ~\citep[e.g., storms and eddies;][]{Showman&Kaspi2013,zhang14,showman19,tan21a,tan21b,tan22}. 
Polarimetric observations \citep{Millar-Blanchaer2020} of the nearby L/T transition, brown dwarf binary system, Luhman\,16AB \citep{luhman13} suggest their atmospheres may be dominated by planetary banding.
Long-term spectrophotometric observations of L/T transition objects have found varying levels of preference for wave models \citep{Apai2017,apai21,Zhou2022,Fuda:2024}, presumably contained within these banded structures. Specifically, \citet{Zhou2022} fit VHS\,J125601.92–125723.9\,b ~\citep[hereafter VHS\,1256\,b, L7,][]{Gauza2015}, a young planetary-mass object, with both wave and spotted models and found a small preference for a 3-wave model over a combination wave/spot model. 

However, Doppler imaging has provided tentative evidence for spot-like features in Luhman 16B through a surface map inferred \citep{crossfield14} using maximum entropy principles. Re-analyzing the same data, \citet{luger21a} identified similar features using the open source \texttt{starry} framework. These results must be interpreted with caution as maximum entropy-based Doppler imaging ~\citep[see e.g.,][]{vogt87} does not deliver a unique solution but instead derives a map in which goodness-of-fit is balanced against the complexity of the image. \citet{crossfield14} further highlighted that as their spectra are sensitive to CO, the inferred maps may be affected by chemical abundance inhomogeneities. Notably, \citet{crossfield14} concluded that zonal bands would not be detectable based on the data precision; therefore, a comparison of spotted versus wave models was not possible at the time. Using \textit{Hubble Space Telescope (HST)} spectrophotometry and the \texttt{Aeolus} code, \citet{karalidi16} identified two pairs of bright and dark features offset by 180$^{\circ}$ on Luhman 16B, features which could fit either spotted or wave paradigms.

SIMP\,J013656.5+093347 ~\citep[hereafter SIMP0136,][]{Artigau2006} is a young ~\citep[200$\pm50$\,Myr,][]{Gagne2017}, L/T transition (T2$\pm0.5$), planetary-mass object ~\citep[12.7$\pm1.0 \ \rm{M_J}$,][]{Gagne2017} demonstrating high variability \citep[$\gtrsim5\%$,][]{Artigau2009,Radigan2014a,Croll2016,Eriksson2019}. Because SIMP0136 is a rapid rotator with a measured period of $2.414\pm0.078\,\rm{h}$ \citep{Yang2016} and $\upsilon \sin i = 52.8\,^{+1.0}_{-1.1}\,\rm{km \ s^{-1}}$ \citep{Vos2017}, surface inhomogeneities \citep[e.g., clouds and storms,][]{Reiners&Basri2008} may explain its history of significant light curve evolution ~\citep[see e.g.,][]{Artigau2009,Apai2013,Apai2017,Croll2016}. Yet, near-infrared (NIR) \textit{Spitzer Space Telescope} observations over 8 epochs and 32 rotations showed SIMP0136's photometry could be best fit by composite sinusoidal functions of wavenumber $k=1,2$; moreover, multiple waves of the same wavenumber with phase offsets can create beating wave patterns, explaining the planetary-mass object's light curve evolution \citep{Apai2017}. \textit{HST} studies of SIMP0136 have detected wavelength-dependent amplitude and phase shifts as well as pressure-dependent contribution functions \citep{Apai2013,Yang2016,McCarthy:2024}, hinting at a complex vertical atmospheric structure. Supporting this interpretation, spectroscopic retrievals \citep{Vos2023} and spectrophotometry \citep{McCarthy:2024} have inferred multiple patchy layers of forsterite and iron clouds in SIMP0136's atmosphere.

In this paper, we analyze multi-band ($Y$/$J$/$H$/$K$) photometry collected for SIMP0136 on two consecutive nights to learn about the source of spectrophotometric variability and atmospheric dynamics. We fit the photometry with both waves and spotted models, assess the performance of each model, and perform a model selection based on goodness-of-fit tests. This leads to insight into SIMP0136's horizontal and vertical atmospheric structure.

The paper is structured as follows. In \S \ref{sec:Observations}, we explain our observational strategy, data reduction methods, and compare the periodic signals of SIMP0136 and our reference stars. We next detail the wave (\S \ref{ssec:models:wave_model}) and storm/spotted (\S \ref{ssec:models:spot_model}) models we use to fit the observed light curves and also explain the metrics we employ to assess model performance (\S\ref{ssec:models:ModelPerformance}). In \S \ref{sec:Results}, we compare our model fits for the photometry of two consecutive nights (\S \ref{ssec:Results:HighCad} and \S \ref{ssec:Results:LowCad}). We discuss the implications for our research in terms of a preferred driver of variability (\S \ref{ssec:discuss:spotsVwaves}), planetary-scale waves' physical nature (\S \ref{ssec:discuss:rossbywaves}), correlation between waves and cloud modulation (\S \ref{ssec:discuss:cloud_formation}), and a detected phase offset between color series (\S \ref{ssec:discuss:phase_shifts}). We summarize our findings and suggest future works in \S \ref{sec:Summary}.

\section{Observations and Data Reduction} \label{sec:Observations} 

NIR photometric observations were conducted at the Canada-France-Hawaii Telescope (CFHT) on the summit of Maunakea, Hawaii using the Wide-field InfraRed Camera (WIRCam; \citealt{Puget2004}). On 14 October 2012 (UT), exclusively $J$-band observations were collected, resulting in high-cadence ($\Delta t = 0.0715$\,h) data. The following night, 15 October 2012 (UT), the filter was alternated between $Y$, $J$, $H$, and $K$ bands, resulting in multi-color but lower cadence ($\Delta t = 0.385$\,h) light curves.

We used the frames reduced by CFHT's default pipeline  (\texttt{iiwi} version 2.1.100; \citealt{Thanjavur2011}) and extracted the photometry from flat-fielded and sky-subtracted frames (``p'' files in the cadc science archive\footnote{\url{https://www.cadc-ccda.hia-iha.nrc-cnrc.gc.ca/en/search/?collection=CFHT&noexec=true}}) using a  fixed 2.8$\arcsec$ aperture radius. The 50$^{\rm th}$ and 90$^{\rm th}$-percentile seeing values were 1.18$\arcsec$ and 1.43$\arcsec$.

The photometric timeseries were obtained using 12 sub-exposures with a per-sub-exposure effective exposure time of 8\,s.  The photometry was measured in individual sub-exposures and averaged to the values used for scientific analysis. As the sequence of sub-exposures is short (${\sim}4$\,min including inter-exposure overheads), we can assume that SIMP0136 and reference stars are stable in flux within the sub-exposure sequence. We therefore use the dispersion of the 12 sub-exposure photometric measurements to determine the photometric uncertainty of their mean value.

\begin{figure*}
\centering
\includegraphics[width=1\textwidth]{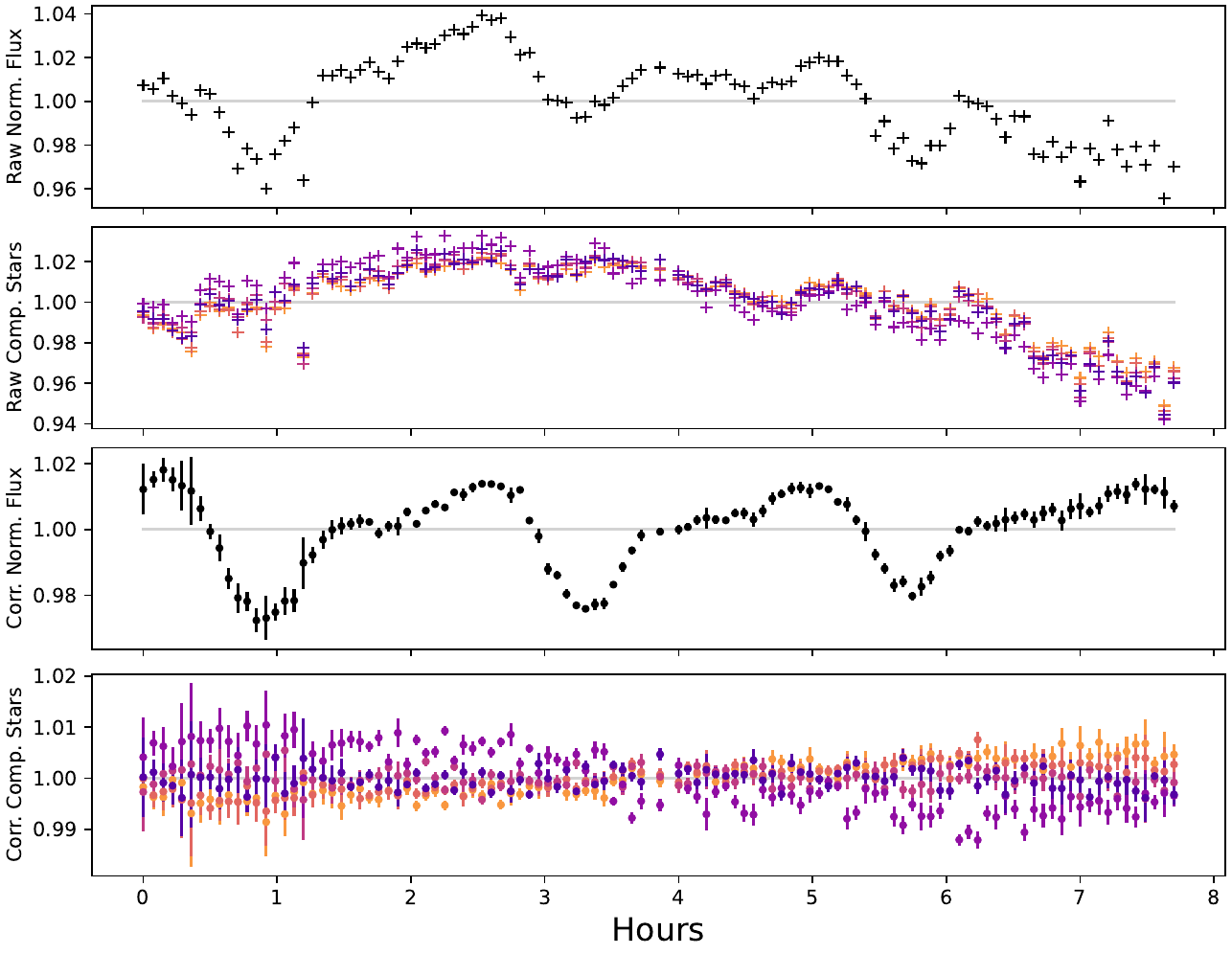}
\caption{\label{fig:J_High_Data_Reduction} High Cadence, $J$-band photometry of SIMP0136 collected at the CFHT on 14 October 2012. (Top) Raw normalized light curves of SIMP0136. (Second Row) Raw normalized light curves of comparison stars. (Third Row) Corrected light curves for SIMP0136. (Bottom) Corrected light curves for comparison stars. Decreased photometric accuracy can be seen at the beginning and end of the observation period due to increased airmass.
}
\end{figure*}

\begin{figure*}
\centering
\includegraphics[width=1\textwidth]{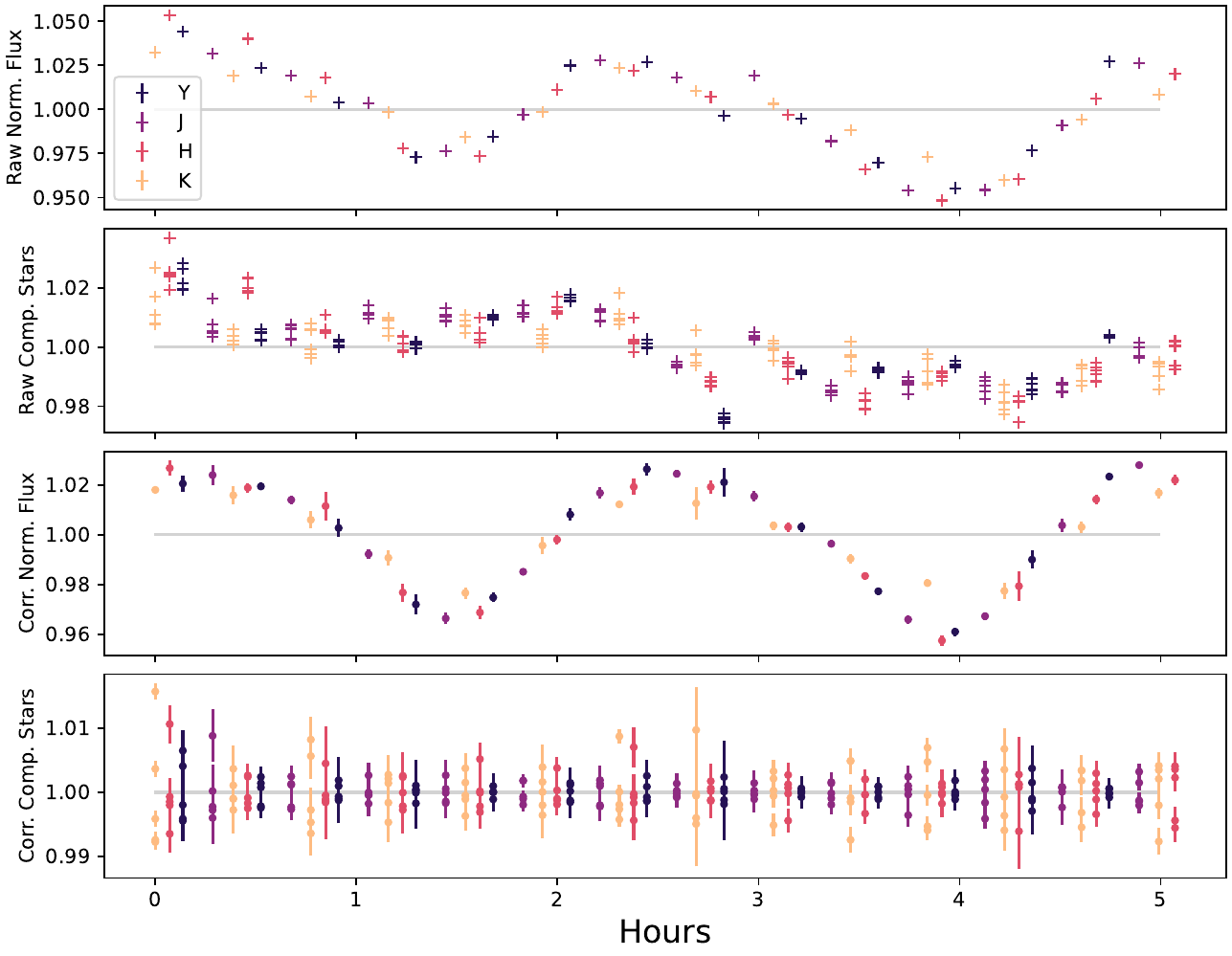}
\caption{\label{fig:Low_Cadence_Data_Reduction} Low Cadence, $Y$/$J$/$H$/$K$ photometry of SIMP0136 collected at the CFHT on 15 October 2012. (Top) Raw normalized light curves of SIMP0136. (Second Row) Raw normalized light curves of comparison stars. (Third Row) Corrected light curves for SIMP0136. (Bottom) Corrected light curves for comparison stars.
}
\end{figure*}

The top two panels of Figures \ref{fig:J_High_Data_Reduction} and \ref{fig:Low_Cadence_Data_Reduction} display the raw (but normalized against each star's mean) light curves for SIMP0136 and 5 reference comparison stars for the 14 October 2012 and 15 October 2012 observations respectively. To correct for systematic effects, SIMP0136 and the comparison stars were divided by the mean normalized light curves of the comparison stars. The corrected light curves can be seen in the bottom two panels of Figures \ref{fig:J_High_Data_Reduction} and \ref{fig:Low_Cadence_Data_Reduction}. 

\begin{figure}
\centering
\includegraphics[width=0.45\textwidth]{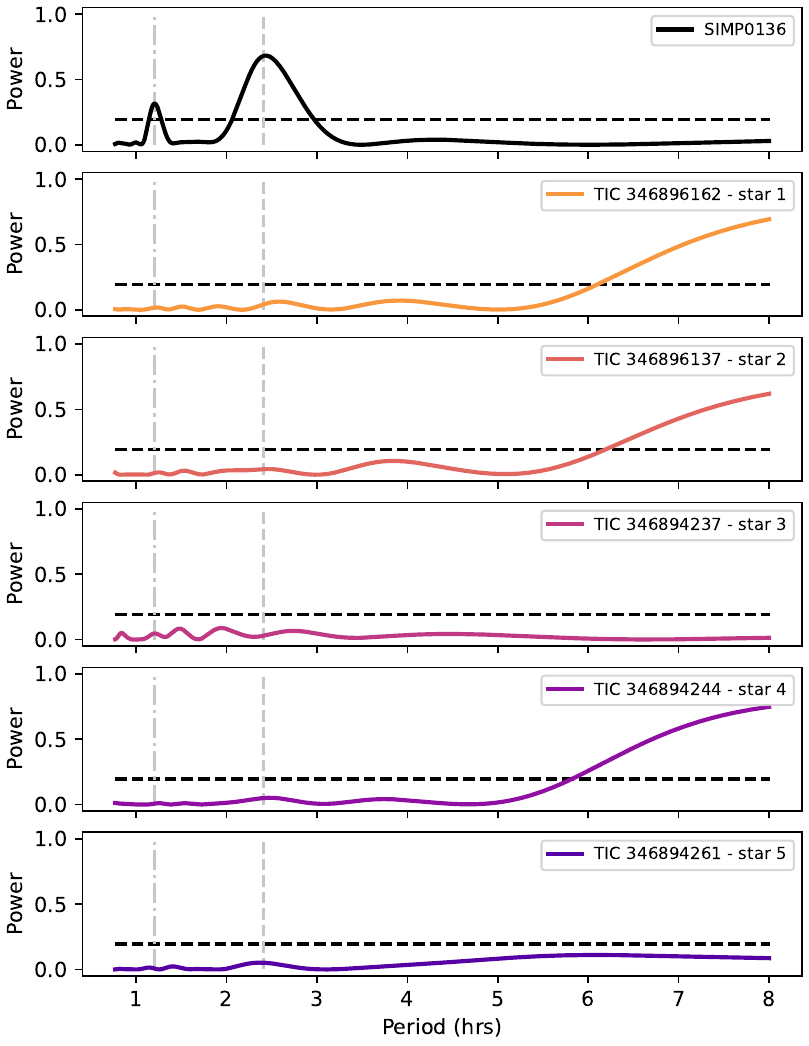}
\caption{\label{fig:Lomb-Scargle} {Lomb-Scargle Periodogram for $J$-band, 14 October 2012 observations. Horizontal, dashed (black) line denotes the 1\% FAP. Vertical, dashed (gray) lines denote the SIMP0136 rotational period (${\sim 2.42}$\,h) and half the rotational period (${\sim 1.21}$\,h). SIMP0136 shows prominent signals matching these periods. Reference comparison stars do not exhibit significant signals $>1\%$ FAP at periodicities near SIMP0136's rotational period.}
}
\end{figure}

The reference comparison stars (see Table \ref{tab:refstars}) were selected due to their location within the NIR detector's field-of-view, brightness, and lack of periodicity near SIMP0136's known rotational rate. As an initial test of of SIMP0136's and the comparison stars' periodicities, we performed Lomb-Scargle (L-S) periodogram analysis \citep{Lomb1976,Scargle1982} using the \texttt{LombScargle} Python class within \texttt{astropy} \citep{astropy:2013,astropy:2018,Astropy2022}. Figure \ref{fig:Lomb-Scargle} shows the results for the $J$-band observations on 14 October 2012. SIMP0136 exhibits two peaks above a 1\% false-alarm-probability (FAP) corresponding to the rotational period (${\sim 2.42}$\, h) found in previous studies ~\citep[e.g.,][]{Artigau2009} and half the rotational period (${\sim 1.21}$\, h). The reference stars do not contain signals above the 1\% FAP near SIMP0136's rotational period. In the appendix, we test the reference stars for trends between airmass and seeing variations and measured brightness. We also confirm the reference stars' non-variability at periods $<1$\,d using external data from the Transiting Exoplanet Survey Satellite ~\citep[TESS;][]{TESS:2015}. 

\begin{table*}
    \centering
    \begin{tabular}{ccccccc}
    \hline
ID &     RA    & DEC  &  $J$ & $H$ & $K$ & TIC \\
 &    GAIA DR3 & GAIA DR3 & 2MASS & 2MASS & 2MASS & \\ \hline
%     24.2796644    & 9.5853710 & 13.252 & 12.809 & 12.585 & 346896171\\ \hline

1&024.27989553291 & +09.5853296147900 & 13.428 & 12.699 & 12.54 & 346896162 \\
2&024.26213598995 & +09.62180319580 & 12.722 & 12.383 & 12.355 & 346896137\\
3&024.16122843640 & +09.55283749631 & 13.762 & 13.451 & 13.366 & 346894237\\
4&024.17202937970 & +09.57556394562 & 13.221 & 12.871 & 12.828 & 346894244\\
5&024.21515175402 & +09.60840662161 & 14.290 & 13.904 & 13.727   & 346894261 \\ \hline
    \end{tabular}
    \caption{Reference comparison stars \citep{Gaia:2023,TESSinputCAT:2018,Cutri2003}.  } 
    % https://ui.adsabs.harvard.edu/abs/2023A%26A...674A...1G/abstract;
    % https://ui.adsabs.harvard.edu/abs/2018AJ....156..102S/abstract
    % https://ui.adsabs.harvard.edu/abs/2003yCat.2246....0C/abstract
    \label{tab:refstars}
\end{table*}

\section{Methods}\label{sec:Models}

To model both planetary-scale atmospheric waves and storms/spots, we modify \texttt{Imber} \citep{Imber,Imber:2024}, an open source Python code. The \texttt{Imber} code, the data used in this work, and a tutorial for duplicating our results and figures are openly available via GitHub\footnote{\url{https://github.com/mkplummer/Imber}} and Zenodo\footnote{\url{https://zenodo.org/doi/10.5281/zenodo.10729261}}. \texttt{Imber} was developed and refined in \citet{Plummer2022,Plummer2023}. A more extensive and complete description of the package and its underlying methodology can be found in those articles.

\texttt{Imber} was created to analytically infer surface inhomogeneities (e.g., magnetic spots, storms, and vortices) on stars, brown dwarfs, and directly imaged exoplanets using a Doppler imaging-based technique for spectroscopic data and light curve inversion for photometry. It allows both data types to be included for an integrated multi-modal solution. \texttt{Imber} also includes a numerical simulation module with a full 3D grid to produce forward models of spectra, spectral line profiles, and light curves. \citet{Plummer2022} demonstrated the numerical and analytical models produce outputs (e.g., line profiles, light curves) with residuals between the two on the order of $0.001$\%. 

This section details the models we employ as well as the metrics by which we evaluate those models. In \S \ref{ssec:models:wave_model}, we describe the wave model we use to fit the photometry. \S \ref{ssec:models:spot_model} provides a brief description of the spotted model we implement via light curve inversion. To infer both wave and spot parameters, we employ Bayesian inference, specifically dynamic nested sampling \citep{Skilling2004,skilling06,higson19} via \texttt{Dynesty} \citep{speagle20} within our framework. The inferred models are evaluated based on fit controlled for the number of free parameters (see \S \ref{ssec:models:ModelPerformance}).

\subsection{Wave Model}\label{ssec:models:wave_model}

For our wave model, we adopt the same approach as \citet{Zhou2022} based on \citet{Apai2017,apai21} and include a bias ($C_0$), linear term ($C_1$), and the sum of multiple ($N$) sinusoidal functions:

\begin{equation}
    F(t) = C_0+C_1 t + \sum_i^N A_i \sin (2 \pi t/P_i + \theta_i).
\end{equation}

The linear term ($C_1$) accounts for a variation on time scales greater than our observation window. $A_i$, $P_i$, and $\theta_i$ are the $i^{\rm th}$ order amplitude, period, and phase. The free parameters are $C_0$, $C_1$, $A_i$, $P_i$, and $\theta_i$. Each additional wave adds three additional free parameters; therefore, the expression for the number of free parameters is thus: $m = 3N+2$.

Uniform priors are assumed for each free parameter in the wave model. Based on SIMP0136's photometric variation of ${\sim}5\%$ in each NIR band, the individual wave components amplitudes ($A_i$) have a range of $2.5 \pm 2.5\%$. For the period, we adjust the range of periods based on the results of the Lomb-Scargle Periodogram in \S \ref{sec:Observations}. Phase varies $180^{\circ}\pm180^{\circ}$. The bias is allowed to vary by $\pm0.25$ and the slope by $\pm0.1\,\rm{h^{-1}}$.

\subsection{Storm/Spot Model}\label{ssec:models:spot_model}

\texttt{Imber} was initially designed to both simulate spotted features' effects on observations and to infer spot parameters given a set of observational data. Here, we will briefly summarize the spotted model, but the authors refer readers to \citet{Plummer2022,Plummer2023} for a more rigorous explanation. 

To enable computationally inexpensive Bayesian inference, the 2D stellar/substellar surface is represented by a 1D flux array. Baseline flux is modeled with a broadening kernel \citep{gray22} with linear limb darkening coefficients selected based on values derived by fitting Luhman\,16B's spectral line profile (due to similar spectral type and $\rm{T_{eff}}$) in \citet{Plummer2022}. Spots are modeled as Gaussian deviations to the baseline kernel to which bright spots add flux and dark spots subtract flux. The added/subtracted flux is scaled by the spot's temperature contrast and apparent size. Size is determined based on the spot radius and also latitudinal and longitudinal foreshortening due to the viewing angle (Lambert's Cosine Law). Object inclination is accounted for with rotation matrices ~\citep[Euler-Rodrigues formula,][]{Shuster1993}. The summed 1D flux at each time step is used to create the modeled light curve.

% Although \citet{Claret2012} limb darkening grids have a minimum temperature of $1500$\,K, these coefficients were demonstrated to adequately fit spectral line profiles of Luhman\,16B ($\rm{T_{eff}}{\sim}1200$\,K) in \citet{Plummer2022}. 

For this work, \texttt{Imber} was modified to allow both radius and temperature contrast to evolve in value from one rotation to the next. The code currently works by setting the rotation by which the evolution is complete, it then varies the spot parameter (radius or contrast) linearly at each time step over one full rotation, thereby, accounting for dynamic atmospheric activity.

The number of spots drives the number of free parameters for these models. Each spot has a latitude, longitude, radius, and temperature contrast. Similar to the wave models, we also include a bias to account for an unknown mean baseline flux. If spot evolution is incorporated into the model, each evolution increases the number of spot parameters. The model leads to the following expression for the number of free parameters: $m = 1+N(4+2E)$, where $N$ is the number of spots and $E$ is the number of spot evolutions.

 % We have also added functionality allowing spots to evolve in size and temperature contrast, potentially adding two free parameters per spot. 

Similar to the wave model, we assume uniform priors for the spotted model free parameters. Latitude and longitude priors encompass the entire sphere ($\pm 90^{\circ}$ and $\pm 180^{\circ}$ respectively). Radius is sampled from $30\pm30^{\circ}$. Contrast varies uniformly from +1 (completely dark spot) to -1 (twice the background brightness).

\subsection{Model Performance Metrics}\label{ssec:models:ModelPerformance} 

% To compare the relative merits of each model, we will use goodness-of-fit tests, and the Bayesian Information Criterion (BIC) to provide metrics. 

% We will start by considering Bayes Rule,

% \begin{equation}\label{eqn:bayesrule}
% P\big(\Theta|D,M\big) = \frac{P\big(D|\Theta,M\big) \ P\big(\Theta|M\big)}{P\big(D|M\big)}
% \end{equation}

% where $\Theta$ is the parameter, $D$ is the data, and $M$ is the model. In traditional Bayesian terms, $P\big(\Theta|D,M\big)$ is the posterior, $P\big(D|\Theta,M\big)$ is the likelihood, $P\big(\Theta|M\big)$ is the prior, and $P\big(D|M\big)$ is the evidence to which we will hereafter refer to as $\mathcal{Z}$ \citep{speagle20}. 

% For this paper we will evaluate models, in part, based on their comparative logarithmic evidence ($\Delta \log \mathcal{Z}$). This provides advantages over more commonly employed Markov-Chain Monte Carlo (MCMC) methods. Generally, MCMC inference focuses on maximizing the posterior, but nested sampling puts an emphasis on maximizing the evidence, providing a natural method for model comparison and normalization \citep{Skilling2004,skilling06}).

To compare the relative merits of each model, we will use goodness-of-fit tests ($\chi^2$ and reduced-$\chi^2$) and the Bayesian Information Criterion (BIC) to provide metrics. Reduced-$\chi^2$ is computed as

\begin{equation}
    \chi^2_\nu = \frac{\chi^2}{\nu} = \frac{\chi^2}{n-m} = \frac{1}{n-m}\sum_{i=1}^{n}\frac{(O_i-M_i)^2}{\sigma^2}
\end{equation}

where $n$ is the number of data points (observations), $m$ is the number of free parameters, $O_i$ and $M_i$ are the $i^{\rm th}$ data points for the observation and model respectively, and $\sigma$ is the photometric uncertainty.

\par We use the following expression to calculate BIC \citep{Kass&Raftery1995}:

\begin{equation}
    \text{BIC} = \chi^2 + m \ln(n) 
\end{equation}

For $\chi^2$, $\chi^2_\nu$, and BIC, a smaller value denotes a better fit. BIC and $\chi^2_\nu$ account for both fit and the number of free parameters, thereby, weighting against more complex models. 

\section{Results}\label{sec:Results}

In this section, we fit the SIMP0136 NIR photometry collected on 14 October 2012 and 15 October 2012 with both wave (see \S \ref{ssec:models:wave_model}) and spotted models (see \S \ref{ssec:models:spot_model}) to determine the primary driver of the planetary-mass object's spectrophotometric variability. Table \ref{tab:Model_Comparison} summarizes the computed $\chi^2_\nu$, $\chi^2$, and BIC values for each model. The preferred model (ranked by $\chi^2_\nu$) is shown at the top of each category; each alternate model has an associated $\Delta \chi^2$ and $\Delta$BIC, denoting its performance with respect to the best fitting model.

\subsection{High-Cadence Data}\label{ssec:Results:HighCad}

We fit the 14 October 2012, high-cadence $J$-band data with both wave and spotted models. For the wave models, we test $N$ = 1 through 3 with improvement seen up to and including 3 waves. We find the 3-wave model is strongly preferred. 4-wave models fail to converge, presumably due to too many free parameters ($m = 14$). We test 1 and 2 spot models, both with and without spot evolution. We attempt 3-Spot models but the models do not converge on a solution. 

% For each model tested, the $\chi^2$, $\chi^2_\nu$, $\Delta\chi^2$, BIC, $\Delta$BIC, with respect to the preferred model can be seen in Table \ref{tab:Model_Comparison}.

\setlength{\tabcolsep}{2pt}
\begin{table*}[t]
\caption{\label{tab:Model_Comparison} Model Comparison: $\chi^2$, BIC, observations ($n$), and free parameters ($m$)}
\centering
    \begin{tabular}{l|c|c|c|c|c|c|c|c|c|c}
    \hline
    \hline
    Observation Date & NIR Band & Model & $\chi^2_{\nu}$ & $\chi^2$ & $\Delta\chi^2$ & BIC & $\Delta$BIC & $m$ \\
    \hline
    14 October 2012 & $J$ & 3-wave & 1.5 & 149 & 0 & 201 & 0  & 11 \\
    & ($n=109$) & 2-wave & 2.4 & 247  & +97.9 & 284 & + 83.8 & 8 \\
    & & 2-spot\,(evolving) & 3.2 & 302 & +154 & 363 & +163 & 13 \\
    & & 2-spot & 3.9 & 389 & +240 & 431 & +230  & 9 \\
    % & & 1-spot (evolving) & 58.9 & 6009  & +5344 & 6042  & +5325  & 7 \\
    % & & 1-spot & 62.5 & 6497 & +5832 & 6520 & +5803 & 5 \\
    \hline
    15 October 2012 & $Y$ & 2-wave & 1.5  & 7.7 & 0 & 29 & 0 & 8 \\
    & ($n=13$) & 1-spot\,(evolving) & 4.4 & 26.1 & +18.4 & 44.1 & +15.9 & 7 \\
    & & 1-spot & 5.7 & 45.3 & +37.6 & 58.1 & +29.9  & 5 \\
    & & 1-wave & 7.3 & 58.3 & +50.6 & 71.1 & +42.9  & 5 \\
    \cline{2-9}
    & $J$ & 2-wave & 2.1 & 10.5 & 0 & 31.0 & 0 & 8 \\
    & ($n=13$) & 1-spot\,(evolving) & 7.0 & 41.9 & +31.4 & 59.9 & +28.8 & 7 \\
    & & 1-wave & 10.4 & 82.8 & +72.3 & 95.6 & +64.6  & 5 \\
    & & 1-spot & 12.2 & 97.6 & +87.1 & 110 & +79.4  & 5 \\
    \cline{2-9}
    & $H$ & 2-wave & 1.8 & 10.5 &  0 & 31.6 & 0 & 8   \\
    & ($n=14$) &  1-wave & 6.8 & 61.5 & +51.0 & 74.7 & +43.1  & 5 \\
    & & 1-spot & 8.6 & 77.2 & +66.7 & 90.4 & +58.8 & 5  \\
    & & 1-spot\,(evolving) & 9.1 & 64.0 & +53.5 & 82.5 & +50.9 & 7  \\
    \cline{2-9}
    & $K$ & 2-wave & 1.0 & 5.75  & 0 & 26.9 & 0 & 8 \\
    & ($n=14$) & 1-wave & 2.6 & 23.5 & +17.8 & 36.7 & +9.8  & 5 \\
    & & 1-spot\,(evolving) & 5.6 & 39.4 & +33.7 & 57.9 & +31.0  & 7 \\
    & & 1-spot & 26.0 & 234 & +228 & 247 & +221  & 5 \\
    \cline{2-9}
    & $Y+J+H+K$ & 2-wave & 11.1 & 509 & 0 & 541 & 0 & 8 \\
    & (n = 54) & 2-spot & 12.8 & 576 & +67 & 612 & +71.0 & 9 \\
    & & 1-spot (evolving) & 13.7 & 642 & +133 & 670 & +129 & 7 \\
     & & 1-spot & 13.8 & 675 & +166 & 695 & +154 & 5 \\
    \hline
    \hline
    \end{tabular}
\end{table*}

\begin{figure*}
\centering
\includegraphics[width=1\textwidth]{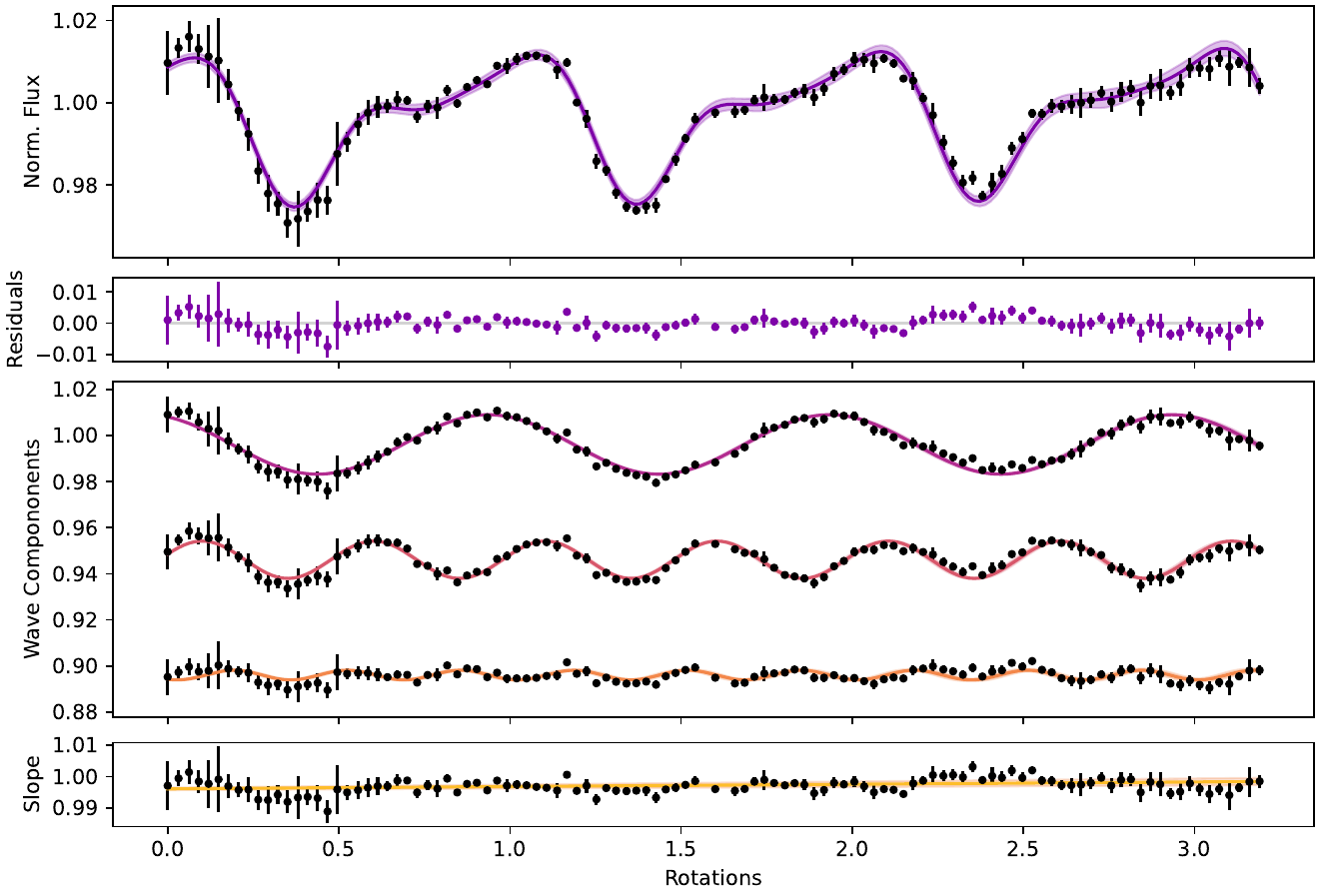}
\caption{\label{fig:J_High_Model&Wave} High cadence, 14 October 2012, $J$-band observations and 3-wave model. Shaded regions are $1\sigma$ uncertainty. Rotations are based on a period of 2.414\,h \citep{Yang2016}. (Top) The inferred 3-wave model fitted to observations. (Second Row) Residuals. (Third Row) 3-wave components with wave numbers ($k$) $\approx$ 1, 2, and 3 and periods of 2.41\, h, 1.21\,h, and 0.80\,h respectively. To demonstrate each component wave's fit, the remaining waves and long-term linear terms are subtracted from the data. (Bottom) Linear term with wave components subtracted from the data.
}
\end{figure*}

Figure \ref{fig:J_High_Model&Wave} shows the preferred 3-wave model with wavenumbers: $k = 1$, $2$, and $3$. Wave amplitude decreases with increasing wavenumber. The $k=1$ period ($P_1 = 2.41\pm0.01$\,h) matches the previously measured rotational period ~\citep[$2.414\pm0.078$\,h,][]{Yang2016}. The $k=2$ and $k=3$ waves' periods ($P_2 = 1.21\pm0.01$\,h and $P_3 = 0.80\pm0.01$\,h) are approximately half and one-third SIMP0136's rotational period. Based on the Nyquist sampling criterion alone, the 14 October 2012, $J$-band mean cadence ($\Delta t = 0.0715$\,h) can theoretically detect signals with wavenumbers $\lesssim 16$, but higher order wave components may be missed due to insufficient signal-to-noise (S/N) ratio. Wave components parameters are summarized in Table \ref{tab:Wave_Parameters}.

We find that SIMP0136 has a $J$-band peak-to-peak variability of $3.85\pm0.14\%$ for 14 October 2012. Here we compute the \% variability based on the minimum and maximum values of the 3-wave model during the observational period. The uncertainty value is computed using the mean of the model standard deviation. The photometric variability measured for 14 October 2012 is lower than that seen the following night, suggesting a dynamic atmosphere over a relatively short time-span.

% The first two waves are in agreement with the L-S periodogram (Figure \ref{fig:Lomb-Scargle}) in \S \ref{ssec:Results:L-S}; however, the L-S periodogram contains a signal peak at ${\sim}1$\,h which is not found in our wave model inference. 

\setlength{\tabcolsep}{2pt}
\begin{table*}[t]
\caption{\label{tab:Wave_Parameters} Inferred Wave Component Parameters: Amplitude ($A_i$), Period ($P_i$), and Phase Shifts ($\theta_i$))}
\centering
    \begin{tabular}{l|c|c|c|c|c}
    \hline
    \hline
    Observation Date & NIR Band & Wavenumber ($k$) & $A_i$\,(\%) & $P_i$\,(h) & $\theta_i$\,(deg) \\
    \hline
    14 October 2012 & $J$ & 1 & $1.29\pm0.02$ &  $2.41\pm0.01$ & $112.43\,^{+2.32}_{-2.33}$ \\
    & & 2 & $0.81\pm0.02$ & $1.21\pm0.01$&  $17.65\,^{+3.69}_{-3.84}$ \\ 
    & & 3 & $0.21\pm0.02$ & $0.80\pm0.01$&  $239.81\,^{+16.71}_{-16.83}$ \\
    \hline
    15 October 2012 & $Y$ & 1 & $2.85\pm0.11$ &  $2.42\pm0.03$ & $58.00\,^{+4.85}_{-5.12}$ \\
    & & 2 & $0.75\pm0.11$ &  $1.27\pm0.02$ & $205.83\,^{+15.54}_{-15.05}$ \\
    \cline{2-6}
    & $J$ & 1 & $2.94\,^{+0.08}_{-0.07}$ &  $2.43\pm0.02$ & $57.95\,^{+3.14}_{-3.23}$ \\
    & & 2 & $0.67\pm0.08$ &  $1.26\pm0.02$ & $190.91\,^{+13.78}_{-12.19}$ \\
    \cline{2-6}
    & $H$ & 1 & $2.86\pm0.09$ &  $2.44\pm0.03$ & $56.93\,^{+4.82}_{-4.94}$ \\
    & & 2 & $0.68\,^{+0.11}_{-0.10}$ &  $1.27\pm0.02$ & $182.53\,^{+15.86}_{-16.58}$   \\
    \cline{2-6}
    & $K$ & 1 & $1.88\pm0.09$ & $2.43\pm0.04$ & $52.29\,^{+7.19}_{-7.06}$   \\
    & & 2 & $0.35\pm0.10$ &  $1.27\pm0.04$ & $151.59\,^{+34.29}_{-32.74}$ \\
    \cline{2-6}
    & $Y+J+H+K$ & 1 & $2.62\pm0.04$ & $2.41\pm0.01$ & $51.95\,^{+2.11}_{-2.10}$\\
    & &  2 &  $0.57\pm0.04$ & $1.26\pm0.01$ & $175.41\,^{+8.44}_{-9.30}$\\
    \hline
    \hline
    \end{tabular}
\end{table*}

The retrieved 3-wave model contains two features that could allow for multi-rotational light curve evolution: offset phases between components and a long-term linear term (slope). The offset phases allow the superposition of each wave to create a dynamic observed light curve.  The Bayesian inference also retrieves a $+0.03\pm0.01\,\% \cdot \rm{h^{-1}}$ linear slope term, demonstrating a gradual increase in flux and hints at dynamics on timescales longer than the period of observation.

Here we briefly highlight a few features from less preferred models. The 2-wave model (second most preferred) retrieves the same period/wavenumber as the 3-wave model's highest amplitude waves: $k=1 \ \rm{and} \ 2$, adding further support to the 3-wave model. The 2-spot model in which spots were allowed to evolve in size and contrast over each rotation returned both a dark spot (radius ${\sim}20^{\circ
}$) and bright spot (radius ${\sim}10^{\circ
}$) and was the third most preferred model. The spots were located at opposite polar latitudes (${\sim} \pm 60^{\circ}$) and varied in both size and contrast over each rotation. The spotted model results will be discussed in further detail in \S \ref{ssec:discuss:spotsVwaves}.

\subsection{Low Cadence Data}\label{ssec:Results:LowCad}

The 15 October 2012 observations provide lower cadence data with significantly fewer data points: 13, 13, 14, and 14 for the $Y$, $J$, $H$, and $K$ bands respectively. The mean cadence between observations of $\Delta t = 0.385$\,h corresponds to maximum detectable wavenumbers $\lesssim 3$ based on the Nyquist sampling criterion. For each NIR band, we are only able to constrain $k=1 \ \rm{and} \ 2$ wave models. 3-wave models do not converge, likely due to the low cadence and small number of observations in each band. 

\begin{figure*}
\centering
\includegraphics[width=1\textwidth]{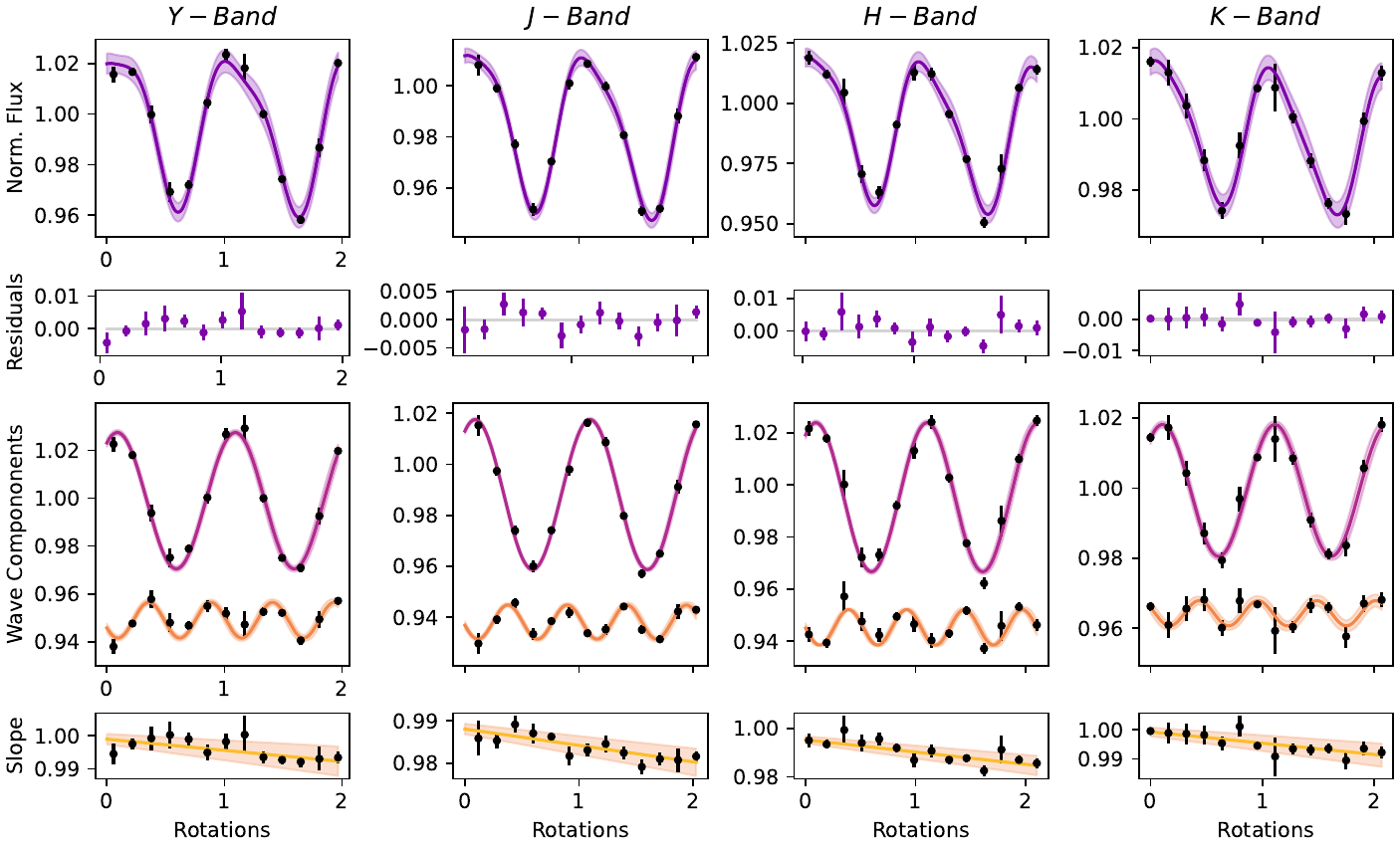}
\caption{\label{fig:Low_Cade_Model&Wave} Low cadence, 15 October 2012, $Y$/$J$/$H$/$K$-band observations and 2-wave models. Shaded regions are $1\sigma$ uncertainty. Rotations are based on a period of 2.414\,h \citep{Yang2016}. (Top Row) Inferred 2-wave model fitted to observations. (Second Row) Residuals. (Third Row, First Column) $Y$-band, components with wave numbers ($k$) $\approx$ 1 and 2 and periods of 2.42\, h and 1.27\,h respectively. (Third Row, Second Column) $J$-band, components with wave numbers ($k$) $\approx$ 1 and 2 and periods of 2.43\, h and 1.26\,h respectively. (Third Row, Third Column) $H$-band, components with wave numbers ($k$) $\approx$ 1 and 2 and periods of 2.44\, h and 1.27\,h respectively. (Third Row, Fourth Column) $K$-band, components with wave numbers ($k$) $\approx$ 1 and 2 and periods of 2.43\, h and 1.27\,h respectively. (Bottom Row) Linear term data and fit. Each wave component and linear term is decomposed as described in Figure \ref{fig:J_High_Model&Wave}.
}
\end{figure*}

Similar to the high cadence, 15 October 2012 data, multi-wave models outperform spotted models for each individual band and when all four NIR bands are fit simultaneously (see Table \ref{tab:Model_Comparison}). The preferred 2-wave models for each NIR band are remarkably similar with each band retrieving $k = 1, 2$ models with periods at the rotational rate and half the rotational rate (see Table \ref{tab:Wave_Parameters} and Figure \ref{fig:Low_Cade_Model&Wave}). The phase offsets also appear similar with each component's $k = 1$ wave having a phase between $52^{\circ}-58^{\circ}$ and $k = 2$ phases ranging between $152^{\circ}-206^{\circ}$ but with overlapping 1$\sigma$ uncertainties. Matching expectations, the simultaneous, composite $Y+J+H+K$ fit retrieves a 2-wave model which is approximately the average of the individual NIR band solutions. 

Using the 2-wave models, we compute peak-to-peak variability of $6.17\pm0.46\%$, $6.45\pm0.33\%$, $6.51\pm0.42\%$, and $4.33\pm0.38\%$ for the $Y$, $J$, $H$, and $K$ bands respectively. Both \% variability and uncertainty are computed as described in \S \ref{ssec:Results:HighCad}. The high variability ($> 5\%$) found here is broadly consistent with previous observations of SIMP0136 \citep{Artigau2009,Apai2013,Apai2017,Metchev2013,Radigan2014a,Wilson2014,Croll2016,McCarthy:2024}.

As can be seen in the bottom panel of Figure \ref{fig:Low_Cade_Model&Wave}, each NIR band retrieval includes a negative linear slope with values of $-0.14\pm0.06\,\% \cdot \rm{h^{-1}}$, $-0.16\pm0.04\,\% \cdot \rm{h^{-1}}$, $-0.21\pm0.05\,\% \cdot \rm{h^{-1}}$ and $-0.16\pm0.05\,\% \cdot \rm{h^{-1}}$ for the $Y$, $J$, $H$, and $K$ bands respectively. This is a steeper slope with the opposite sign as that seen in the 14 October 2012 data the night prior. The variation may lend evidence to unmodeled waves or other dynamics with timescales greater than the period of observation.

Spotted models have greater difficulty modelling the low cadence data. Evolving 2-spot models require approximately an equal number of free parameters as there are data points (and are therefore not considered) while non-evolving, 2-spot models have difficulty converging (with the exception of the composite $Y+J+H+K$ fit which has the benefit of a higher number of observations). Both evolving and non-evolving 1-spot models retrieve polar spots (${\sim}70^{\circ}-80^{\circ}$) with radii of ${\sim}30^{\circ}$. Evolving 1-spot models tend to outperform non-evolving models, with the exception being the $H$ band where spot evolution is preferred by BIC but not $\chi^2_\nu$. 

% and outperform 1-wave models in the $Y$ and $J$ bands.

% The $J$ and $K$ bands each recover large high-latitude dark and bright spot pairs near opposite poles. The $H$-band returns a similarly sized dark/bright spot pair at mid-latitudes. If SIMP0136 truly possesses spots of this nature, these results would point to heterogeneity in the planetary-mass object's vertical atmospheric structure, supporting layered cloud models of various thicknesses \citep{Apai2013,Vos2023}.

% \subsection{Multi-Epoch $J$-Band Fitting}

% Because SIMP0136, and similar L/T transition objects, demonstrate multi-rotational light curve evolution, we take advantage of the consecutive nights of observation to explore this phenomenon. We simultaneously fit both observational periods with wave and spotted models, allowing the spotted models to evolve in size and temperature contrast between the nights.

\section{Discussion}\label{sec:Discussion}

\subsection{Storms or Waves as Primary Variability Driver?}\label{ssec:discuss:spotsVwaves}

As can be seen in Table \ref{tab:Model_Comparison}, wave models outperform spotted models in terms of $\chi^2$, $\chi^2_\nu$, and BIC. Here we seek to explore physical explanations for these results.

To further understand SIMP0136's atmospheric dynamics, we will use two planetary-scale parameters: the Rossby deformation radius ~\citep[$L_D$, e.g.,][]{Gill1982} and the Rhines scale ~\citep[$L_{\beta}$,][]{Rhines1975}. The Rossby deformation radius is the length at which rotational (Coriolis) effects become important \citep{Gill1982}, and it can also be seen as the typical scale for atmospheric storms and vortices  ~\citep[e.g.,][]{tan21b,Zhou2022}. The Rossby deformation radius is computed (in km) as follows \citep{Showman2013}:

\begin{equation}
    L_D = \frac{U}{2 \Omega \sin{\phi}},
\end{equation}
where $U$ is the flow speed, $\Omega$ is the angular rotational speed, and $\phi$ is latitude.

At lengths greater than the Rhines scale, atmospheric structures transition from turbulent features to zonal jets as seen on Solar System planets ~\citep[e.g.,][]{Cho&Polvani1996,Showman2010,Showman2013,Haqq-Misra2018}; here the Rhines scale is computed (in km):

\begin{equation}
    L_{\beta} = \sqrt{\frac{UR}{2 \Omega \cos{\phi}}},
\end{equation}

where $R$ is the object's radius

Both inferred spots exceed the Rossby deformation radius and Rhines scale for SIMP0136, meaning the retrieved spotted models are likely unphysical. Considering the spotted model with the best $\chi^2_{\nu}$ (see Table \ref{tab:Model_Comparison}), the high-cadence $J$-band, 2-spot model with size and contrast evolution, we compute the Rossby deformation radius and Rhines scale at the inferred spot latitude. We conservatively assume a flow velocity ($U$) of $1000\,\rm{m\,s^{-1}}$ based on the brown dwarf \hbox{2MASS J10475385+2124234}'s \citep{Burgasser1999} measured wind speed of $650\,\rm{m\,s^{-1}}$ \citep{Allers2020}. We assume a planetary radius of 1.15\,$R_J$ based on spectral energy distribution analysis of SIMP0136 by \citet{Vos2023}. These assumptions result in $L_D = 0.9^{\circ}$ and $L_{\beta} = 11.8^{\circ}$ for latitudes of $\pm60^{\circ}$. The dark spot at latitude $-62.36^{\circ}\,^{+1.37^{\circ}}_{-1.23^{\circ}}$ has inferred radii varying from $23.68^{\circ}\,^{+4.67^{\circ}}_{-3.00^{\circ}}$ to $20.27^{\circ}\,^{+5.77^{\circ}}_{-3.31^{\circ}}$. The bright spot has a latitude of $+56.77^{\circ}\,^{+6.28^{\circ}}_{-19.04^{\circ}}$ and radii ranging from $11.20^{\circ}\,^{+3.24^{\circ}}_{-2.33^{\circ}}$ to $10.78^{\circ}\,^{+2.73^{\circ}}_{-1.99^{\circ}}$. We can see the dark spot's inferred radius range exceeds $L_{\beta}$ and both spots' radii are far greater than $L_D$. It is also unlikely that a group of adjacent spots of equivalent summed size are responsible for the observed variability as the same arguments above would require such spots to be in separate latitudinal bands (based on the Rhines scale) in which differential rotation would disperse coherently-structured spot groupings.

The retrieved spots' polar latitudes also argue against a spotted model explaining SIMP0136's variability. When exploring different inclinations through GCMs, objects viewed equator-on generate higher variability light curves than those viewed pole-on \citep{tan21b}. For rotational periods on the scale of SIMP0136 (${\sim}2.5$\,h), GCMs also demonstrate that equatorial regions have more enhanced temperature variation and cloud vertical extent than polar latitudes. This is in alignment with observations indicating that brown dwarfs' equatorial regions are more variable and redder \citep{Vos2017,Vos2018,Vos2020,Vos2022} and also cloudier \citep{Suarez2023} than higher latitudes. This argument implies that polar structures similar to Saturn's polar hexagonal feature \citep{Godfrey:1988} or Jupiter's circumpolar cyclones \citep{Bolton:2017,Adriani2018,Orton:2017} are also unlikely to be the dominant driver of SIMP0136's variability.

% The Rossby deformation radius and Rhines scale divide rotating objects such as planets and brown dwarfs into three atmospheric regimes: slow rotators, Rhines rotators, and rapid rotators \citep{Haqq-Misra2018}. For slow rotators, both the Rossby deformation radius and the Rhines scale are greater than the planetary radius, leading to large-scale turbulent features \citep{Haqq-Misra2018}. Rhines rotators are objects in which the Rhines scales exceed the planetary radius but the Rossby deformation radius does not; these objects may potentially exhibit zonal banding at lower latitudes with turbulent features expressing at mid-latitudes \citep{Haqq-Misra2018}. 

% Rapid rotators are objects in which both parameters are less than the planetary radius and therefore planetary banding is the primary driver of atmospheric features \citep{Haqq-Misra2018}. Because of its high rotation rate, SIMP0136 is considered a rapid rotator and its large-scale atmospheric dynamical structure is expected to be dominated by planetary banding. 

\subsection{Planetary-scale Waves in Substellar Atmospheres}\label{ssec:discuss:rossbywaves}

The preference for atmospheric waves found in this work is in agreement with prior studies of SIMP0136 photometry \citep{Apai2017,McCarthy:2024} as well as other L/T transition dwarfs ~\citep[e.g.,][]{apai21,Zhou2022,Fuda:2024}. Similar to our results in \S \ref{ssec:Results:HighCad} and \S \ref{ssec:Results:LowCad}, peak signals corresponding to $k = 1$ and $k=2$ waves have been previously identified for SIMP0136 \citep{Apai2017}. {Based on observations over hundreds of rotations and thousands of hours, Luhman\,16B has been found to also contain signals corresponding to groups of $k = 1,2$ waves with three to four sine waves accounting for long-term light curve variability \citep{apai21,Fuda:2024}. These results suggest latitudinal variation in wind speeds and atmospheric waves within a banded structure \citep{Fuda:2024}}. For VHS\,1256\,b, a 3-wave model was found to best fit observations over 2 rotations \citep{Zhou2022}. The 3-wave model was comprised of two waves ($P_1,P_2 =18.8\pm0.2\,\rm{h}, 15.1\pm0.2\,h$), less than the rotational period ~\citep[$22.02\pm{0.04}\,\rm{h}$,][]{Zhou2020b}, forming a beating pattern and a third, $k=2$ wave with $P_3 = 10.6\pm0.1\,\rm{h}$ \citep{Zhou2022}.

Brown dwarf 3D GCMs also support atmospheric waves driving variability at equatorial latitudes. For rapid rotators like SIMP0136, GCMs exhibit equatorial waves with longer zonal wavelengths and lower wavenumber values (similar to those found in \S \ref{ssec:Results:HighCad} and \ref{ssec:Results:LowCad}), as well as the enhanced cloud coverage and temperature variation in their equatorial regions discussed in \S \ref{ssec:discuss:spotsVwaves} \citep{tan21b}. Strong evidence is also found for cloud radiative feedback-driven Kelvin waves (and more tentative evidence for Rossby waves) zonally propagating at equatorial latitudes, contributing to light curve variability \citep{tan21b}. Kelvin waves move along a barrier, which in this case is formed by the equator (along which, the wave moves eastward); essentially, the forces pushing the fluid pole-ward due to a pressure gradient are balanced by Coriolis forces acting towards the equator in eastward moving fluids ~\citep[e.g.,][]{Gill1982}. Rossby waves, an additional species of large-scale atmospheric wave, are driven by the conservation of potential vorticity (a fluid mechanics analog to the conservation of angular momentum) and the variation of the Coriolis parameter with latitude \citep{Rossby1945}. 

Planetary-scale waves have been observed to play an important role in the atmospheric dynamics of Jupiter. Quasi-stationary and alternating patterns of relatively cloud-free NIR hot-spots and cooler ammonia cloud-enhanced plumes have long been observed to be associated with the jet at the boundary of Jupiter's North Equatorial Belt (NEB) and Equatorial Zone (EZ) \citep{Choi2013}. These features are widely considered to be driven by a Rossby wave within Jupiter's equatorial region \citep{Allison1990,Showman2000,Friedson2005} with the wave crests correlating to the ammonia aerosol-enhanced plumes and troughs with cloud-free regions due to condensate sublimation \citep{dePater:2016,Fletcher:2016,Fletcher2020}.

% Rossby waves have been shown to primarily modulate the ammonia aerosol content within hot spots and plumes with these features showing neither elevated nor diminished temperatures or ammonia abundances compared to their surroundings \citep{Fletcher2020}.

\subsection{Cloud Modulation and Breakup Associated with Atmospheric Waves}\label{ssec:discuss:cloud_formation}

Multi-band photometry offers the opportunity to explore if cloud modulation associated with planetary-scale waves (as seen in brown dwarf GCMs and observations of Jupiter) exists in SIMP0136's weather layer. If the variability from the waves is associated with silicate and metal clouds \citep{Suarez&Metchev2022,Vos2023,McCarthy:2024}, the light curve minima are where we would expect cloud coverage. In this scenario, the light curve maxima would be associated with hot spots, atmospheric areas of depleted aerosols in which light from deeper atmospheric layers is observable. We expect clouds to scatter and therefore redden light, and indeed, cloudier brown dwarfs have been found to be redder than less cloudy objects \citep{Suarez&Metchev2022}. Light curve minima should be redder and maxima bluer if this hypothesis is true. A lack of correlation between the light curves and NIR color might support an alternate theory such as convective fingering \citep{Tremblin2016,Tremblin2019}.

{To date, conclusive color variability has not been detected in SIMP0136 or similar L/T transition objects. \citet{Artigau2009} reported correlated $J$ and $K_S$ band light curves ($\frac{\Delta K_S}{\Delta J} = 0.46 \pm 0.06$) for SIMP0136 which they found to best fit a scenario with dusty clouds in a clear atmosphere with two temperature components. \citet{Radigan2012} recorded comparable results ($\frac{\Delta H}{\Delta J}$ and $\frac{\Delta K_S}{\Delta J}$) for 2MASS\,J21392676+0220226 ~\citep[2M2139, T1.5,][]{Reid2008} and found the data suggested these ratios may be variable, particularly $\frac{\Delta K_S}{\Delta J}$. Using HST data, \citet{Apai2013} detected shallow $J-H$ color variability for both SIMP0136 and 2M2139. \citet{Lew2020a} found similarly negligible $J-H$ variability across a population of L/T transition substellar objects including SIMP0136. Recently, a $J-K_S = 0.03$ color change was reported by \citet{McCarthy:2024} using NIR photometry collected with the 1.8\,m Perkins Telescope Observatory, perhaps indicating that redder wavelengths should be considered for color variability studies.}

\begin{figure*}
\centering
\includegraphics[width=1\textwidth]{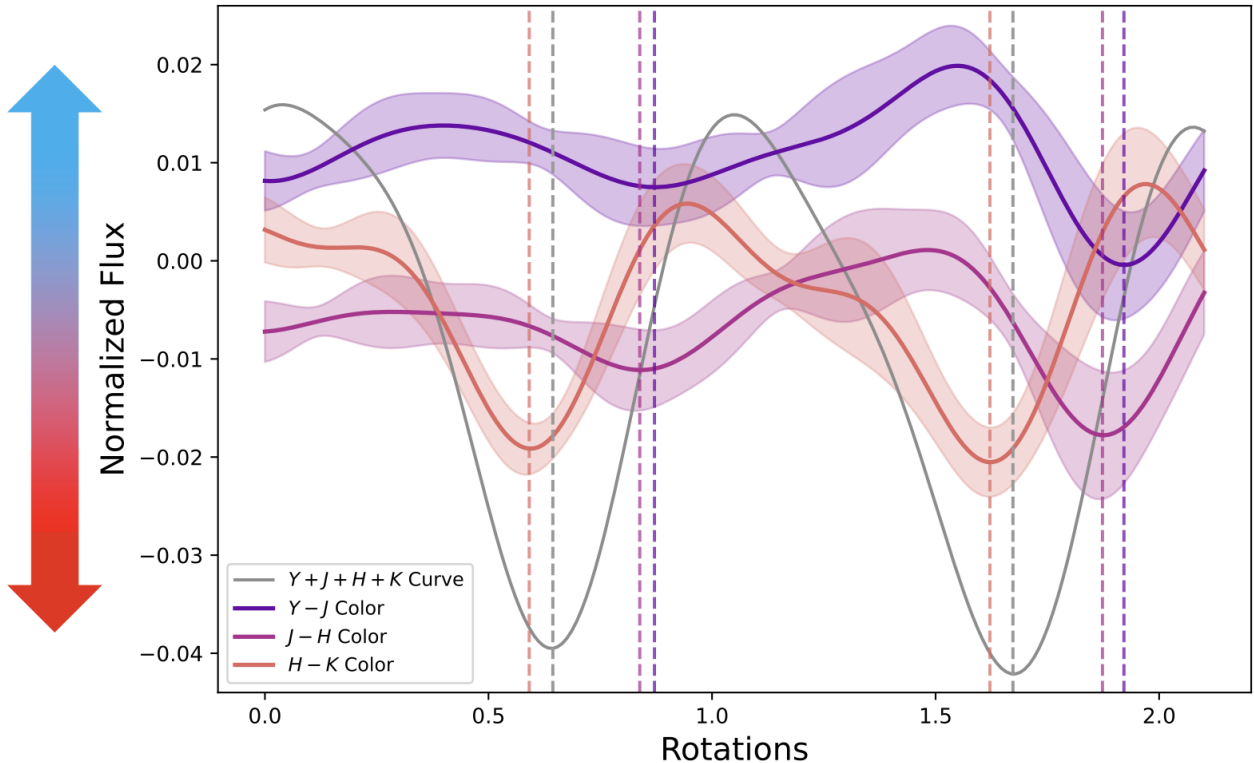}
\caption{\label{fig:Color_Comparison} Color time series comparison to composite $Y+J+H+K$ light curve (gray). Shaded regions are $1\sigma$ uncertainty. Vertical dashed lines denote color series and light curve minima. Qualitatively, it can be seen that the composite light curve minima correspond to reddening (i.e. low flux in $Y-J$, $J-H$, and $H-K$ corresponds to redder color indices). This provides evidence for cloud scattering (see \S \ref{ssec:discuss:cloud_formation}). Light curve maxima correspond to bluer, presumably relatively cloud-free regions. The $Y-J$ and $J-H$ color series are approximately in phase but the $H-K$ solution is ${\sim}90^{\circ}$ out of phase with the remaining color series, offering evidence of SIMP0136's complex vertical cloud structure.
}
\end{figure*}

Using the $Y$/$J$/$H$/$K$, 15 October 2012 photometry, we conduct a preliminary analysis to determine if the light curve minima are associated with redder NIR colors and if light curve maxima are associated with corresponding bluer colors. As a proxy for color, we use the normalized flux of the 2-wave models to create $Y-J$, $J-H$, and $H-K$ color time series (see Figure \ref{fig:Color_Comparison}). Qualitatively it can be seen that for each color series, the light curve (here we use the simultaneously-inferred composite $Y+J+H+K$ curve) minima approximately correspond to significant dips towards redder colors while maxima correspond to bluer colors. This behaviour is more pronounced for longer wavelengths (e.g., $H-K$). The $H-K$ color time series leads the composite $Y+J+H+K$ curve to a small degree while $Y-J$ and $J-H$ each lag the light curve minima. We interpret these results as tentative evidence for complex vertical cloud structures with (presumably silicates and metal) clouds located at the minima and cloud-free regions existing at the maxima due to condensate sublimation. 

Follow-up observations using, ideally space-based, platforms with NIR and mid-IR (MIR) spectroscopic capabilities such as the \textit{James Webb Space Telescope (JWST)} would have the capability to gather broad wavelength times series (see e.g., \textit{JWST} Cycle 2 GO Program 3548, PI J. Vos). With this data, broadband flux variations could be compared to  variations in cloud coverage and chemical abundances, further constraining the nature of planetary-scale waves.

\subsection{Phase Shifts Between NIR Bands}\label{ssec:discuss:phase_shifts}

As wavelengths are sensitive to different pressure levels in brown dwarf and planetary atmospheres ~\citep[e.g., Figure 4 in][]{McCarthy:2024}, phase shifts between NIR bands can indicate inhomogeneity in their vertical structures. {Space-based observations over baselines ranging between ${\sim}5{\textendash}50$\,h detected large phase shifts (${\sim}90{\textendash}180^{\circ}$) between shorter NIR bands (e.g., $J$ and $K$) and longer wavelengths (e.g., \textit{Spitzer}/Infrared Array Camera Ch.1 [$3.6\,\mu m$] and Ch.2 [$4.5\,\mu m$]) \citep{Buenzli2012,Yang2016} for 2MASS J22282889{\textendash}4310262 ~\citep[T6.5,][]{Burgasser2003}, 2MASS J15074769{\textendash}1627386 ~\citep[L5,][]{Reid2000}, and 2MASS J18212815+1414010 ~\citep[L5,][]{Looper2008}. }

L/T transition objects such as VHS\,1256\,b, Luhman\,16B (T0.5), 2M2139, and SIMP0136 have demonstrated, in general, subtler phase shifts than earlier L and later T dwarfs. \citet{Zhou2022} collected 42\,h of time series observations of VHS\,1256\,b using the \textit{HST}/Wide Field Camera 3 (WFC3) and identified no discernible phase shift between the F127M ($1.270\pm0.035\,\rm{\mu m}$), F139M ($1.385\pm0.035\,\rm{\mu m}$), and F153M ($1.530\pm0.035\,\rm{\mu m}$) filters. {Observing Luhman\,16B with a 2.2\,m ground-based telescope over 4\,h in optical and NIR bands, \citet{biller13} detected significant phase offsets between $K$-band light curves and both $H$ and $z'$ bands. However, observing the same object over 6.5\,h with \textit{HST}/WFC3, \citet{Buenzli15a} found the $J$, $H$, and water bands to all be in phase. \citet{Apai2013} did not find phase shifts in either 2M2139 or SIMP0136 using \textit{HST}/WFC3 G141 data over baselines of ${\sim5.9}$\,h and ${\sim3.1}$\,h respectively, but \citet{Yang2016} found modest phase shifts (${\sim}30^{\circ}$) between light curves derived from \textit{HST}/WFC3 G141 and \textit{Spitzer} Ch.1 and Ch.2 over baselines of ${\sim}10$\,h. \citet{McCarthy:2024} similarly found phase shifts between $J$ and $K_s$ bands of $39.9^{\circ}\,^{+3.6^{\circ}}_{-1.1^{\circ}}$ using a 1.8\,m ground-based telescope over a 8.5\,h baseline.}

% For the SIMP0136 observations presented in this work, when possible, we compare the phases of the inferred wave components (e.g. $k = 1, 2, 3$) between the $J$, $H$, and $K$ bands as well as the phase shift between each band's wave components.

Considering the color time series in Figure \ref{fig:Color_Comparison}, it can be qualitatively seen that the $H-K$ series is offset by ${\sim}90^{\circ}$ from the $J-H$ and $Y-J$ color series, which are approximately in phase with one another. Adopting a similar approach as \citet{McCarthy:2024}, we use the \texttt{signal} function within \texttt{Scipy} \citep{scipy2020} to determine the phase shift via cross-correlation. Performing cross-correlation on 5,000 samples from the $H-K$ and $J-H$ solutions provides a phase shift of $93.8^{\circ}\pm7.4^{\circ}$. This $12.7\sigma$ detection provides evidence for complex vertical cloud structure between the pressure levels corresponding to these NIR bands, likely in the form of multiple cloud layers as suggested by \citet{Vos2023} and \citet{McCarthy:2024}.

Statistically significant phase shifts between the inferred NIR band wave components are not detected for SIMP0136 in this work (see Table \ref{tab:Wave_Parameters}). Within each band, the $k = 1$ and $k = 2$ components are offset by approximately $100^{\circ}-140^{\circ}$ from one another. The $k = 1$ components each have phases ranging from $\sim52^{\circ}-58^{\circ}$. The $k = 2$ components have phases ranging from $\sim150^{\circ}$ to $205^{\circ}$. For both $k=1$ and $k=2$ waves, the phase shifts between NIR bands are less than the $3\sigma$ uncertainty.

% The 15 October 2012 observations include near-simultaneous observations in $Y$, $J$, $H$, and $K$ bands, so we can compare phase shifts between both NIR bands and wave components. 

Higher cadence observations in both the NIR and MIR, such as those capable by the \textit{JWST}, would be able to reduce phase uncertainty and detect phase shifts at lower pressure levels (higher altitudes). Such detections would provide further collaborating evidence of cloud modulation associated with planetary-scale waves along with a more complete picture of SIMP0136's vertical architecture.    

% For the 14 October 2012 $J$-band observations, we find a phase shift between the $k=1$ and $k = 2$ components of ${\sim}90^{\circ}$. The $k = 3$ phase has a high uncertainty ($\pm179^{\circ}$); therefore, this component is not compared to the others.

% The lack of significant phase shift detection between $J$, $H$, and $K$ bands is consistent with previous findings for L/T transition dwarfs, as detailed above. 

% The detected phases shifts are indicative of inhomogeneity, modulated by the $k=2$ wave component, in SIMP0136's vertical structure at the pressures probed by these wavelengths. These phase shifts provide further collaborating evidence of cloud formation associated with planetary-scale waves. 

% Observations in the MIR, such as those capable by the \textit{JWST}, would be able to detect phase shifts at lower pressure levels (higher altitudes) and provide a more complete picture of SIMP0136's vertical architecture. 

\section{Summary}\label{sec:Summary}

To determine the driving source of spectrophotometric variability in SIMP0136, an L/T transition object, we analyze NIR photometry collected at the CFHT on the nights of 14 October 2012 and 15 October 2012. The data provides coverage for ${\gtrsim}5$ rotations of SIMP0136. We modify the publicly available, open source Python code \texttt{Imber} \citep{Imber,Imber:2024}, developed and honed in \citet{Plummer2022,Plummer2023}, to fit the observed light curves with waves as well as spotted models. Here are our findings:

\begin{enumerate}
    \item The 14 October 2012, high cadence $J$-band observations are best fit with a 3-wave model consisting of $k = 1, 2, \ \rm{and} \ 3$ components with periods 2.41\,h, 1.21\,h, and 0.80\,h (see \S \ref{ssec:Results:HighCad}). A linear term with a positive slope is also retrieved and indicates an increase in flux throughout observation.

    \item The 15 October, low cadence $Y$/$J$/$H$/$K$ light curves are each fit with 2-wave models with $k = 1 \ \rm{and} \ 2$ components (see \S \ref{ssec:Results:LowCad}). Each of these components has periods approximating the rotational period and half the rotational period. In each of the inferred models, the linear term has a negative slope, demonstrating a change from the prior night's observations.

    \item For the spotted models, the retrieved spot radii exceed the Rossby deformation radius (and Rhines scale for the dark spot) at the inferred latitudes and assumed mean flow speed and planetary radius, indicating such spots to likely be unphysical (see \S \ref{ssec:discuss:spotsVwaves}).

    \item For the multi-band ($Y$/$J$/$H$/$K$) photometry, the inferred 2-wave models demonstrate correlation with shifts in color (see \S \ref{ssec:discuss:cloud_formation}). The light curve minima appear to correspond to redder colors while the maxima correspond to bluer flux. This correlation is strongest for the $H-K$ color time series. We propose this may be tentative evidence for planetary-scale waves traveling in the vertical plane of SIMP0136's atmosphere with enhanced silicate or iron cloud coverage in the wave crests and depleted aerosols in the troughs.

    \item  We detect a $93.8^{\circ}\pm7.4^{\circ}$ ($12.7\sigma$) phase offset between the $H-K$ and $J-H$ color series, providing evidence for complex vertical cloud structure in SIMP0136's atmosphere (see \S \ref{ssec:discuss:phase_shifts}).

\end{enumerate}

Moving forward, a greater understanding of the true nature of planetary-scale waves potentially driving brown dwarf and planetary-mass object variability can be achieved by multi-rotational and quasi-simultaneous observations in the NIR and MIR by platforms such as \textit{JWST}. Spectroscopic modes would help to discern if the reddening associated with NIR light curve minima is tied to variations in cloud coverage or chemical abundances. Both photometric and spectroscopic modes could further constrain phase shifts between NIR and MIR bands and color indices, providing a greater understanding of SIMP0136's vertical structure throughout its atmospheric layers. 

% \begin{table}[t]
% \caption{\label{tab:InstrComp} Inferred Spot Evolution}
% \centering
%     \begin{tabular}{c |c| c| c}
%     & & Dark Spot & Bright Spot \\
%     \hline
%     Spot & Initial & 27.9 & 16.3 \\
%     Radius & Final & 21.2 & 16.4 \\
%     (deg) & $\Delta$ & -6.7 & +0.1 \\
%     \hline
%     Spot  & Initial & 678 & 1756 \\
%     Temperature & Final & 705 & 1742 \\
%      (K) & $\Delta$ & +27 & -14 \\
%     \hline
%     \end{tabular}
% \label{tab:datasets}
% \end{table}

\section*{Acknowledgements}

\par M.K.P. would like to thank the United States Air Force Academy's Department of Physics and Meteorology, the United States Air Force Institute of Technology's Civilian Institution Program, and The Ohio State University's Department of Astronomy for supporting and enabling this research. J.W. acknowledges the support by the National Science Foundation under Grant No. 2143400. Additionally, we want to acknowledge the hard work and expertise of the scientific, technical, and administrative staff at the Canada-France-Hawaii Telescope. We thank Leigh N. Fletcher for his professional insight on solar system atmospheres. {We thank the anonymous reviewer for their constructive feedback and comments.} We would also like to thank the Group for Studies of Exoplanets (GFORSE) at The Ohio State University, Department of Astronomy for continuous feedback throughout the development of this research. 

\par The results presented in this paper are based on observations obtained at the Canada-France-Hawaii Telescope (CFHT) which is operated by the National Research Council (NRC) of Canada, the Institut National des Sciences de l'Univers of the Centre National de la Recherche Scientifique (CNRS) of France, and the University of Hawaii. Based on observations obtained with WIRCam, a joint project of CFHT, Taiwan, Korea, Canada, France, at the CFHT. The observations at the CFHT were performed with care and respect from the summit of Maunakea which is a significant cultural and historic site.

\par This paper includes data collected by the TESS mission. Funding for the TESS mission is provided by the NASA's Science Mission Directorate.

\par The views expressed in this article are those of the author and do not necessarily reflect the official policy or position of the Air Force, the Department of Defense, or the U.S. Government.

\software{Imber \citep{Imber,Imber:2024}, Astropy \citep{astropy:2013, astropy:2018,Astropy2022}, Dynesty \citep{speagle20}, Lightkurve \citep{Lightkurve:2018}, Matplotlib \citep{Matplotlib}, Pandas \citep{Pandas}, Scipy \citep{scipy2020}}

\appendix
\section{Reference stars}\label{appendix_A}
{In our analysis above, we assumed the reference stars are sufficiently stable on timescales commensurate with the rotation period of SIMP0136 to serve as photometric calibrators. Reference stars were chosen based on: their location on the same WIRCam science array as SIMP0136, their brightness ($J<12.8$), and the need for these stars' signals to be unsaturated and unaffected by known bad pixels. Coordinates and $JHK$ magnitudes of the reference stars used in our photometric analysis are listed in Table~\ref{tab:refstars}. Four stars have closely matching colors ($J-K$ ranging from 0.367 to 0.396) while star \#1 has a significantly redder $J-H=0.89$. }

\begin{figure}
\centering
\includegraphics[width=0.8\textwidth]{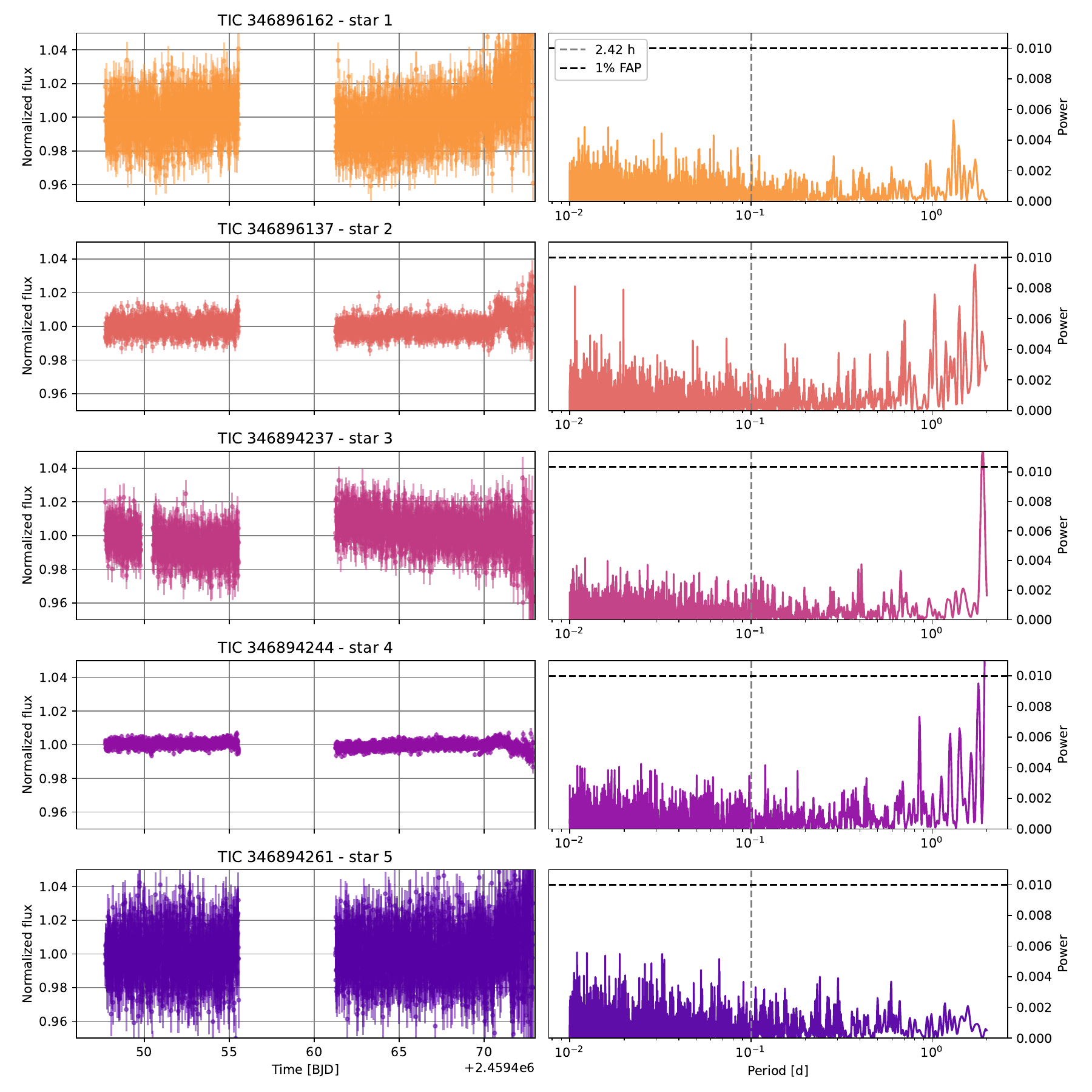}
\caption{TESS photometric lightcurve for the 5 reference stars (left). None of the stars display short-term ($<1$\,day) variability above a 1\% false alarm probability and no signal is seen close to the rotation period of SIMP0136. \label{fig:ref_stars_lc}}.
\end{figure}

{While there is no simultaneous external calibration of the photometric stability of our reference stars, the TESS satellite provides month-long observations of nearly the entire sky. The SIMP0136 field has been visited once by TESS (Sector 42, August-September 2021), providing timeseries of our reference stars. SIMP0136 is too faint for useful monitoring with TESS due to its extremely red optical-to-infrared colors, but the five reference stars are sufficiently bright for useful photometry to be retrieved. We extracted their background-subtracted lightcurves with the \textsc{lightkurve} package \citep{Lightkurve:2018}. These observations provide insight into the variability of our reference stars. As shown in Figure~\ref{fig:ref_stars_lc}, none of the stars display short-timescale percent-level variability, and they do not display significant periodicity above the 1\% FAP level for periods shorter than 1 day.}

\begin{figure}
\centering
\includegraphics[width=0.8\textwidth]{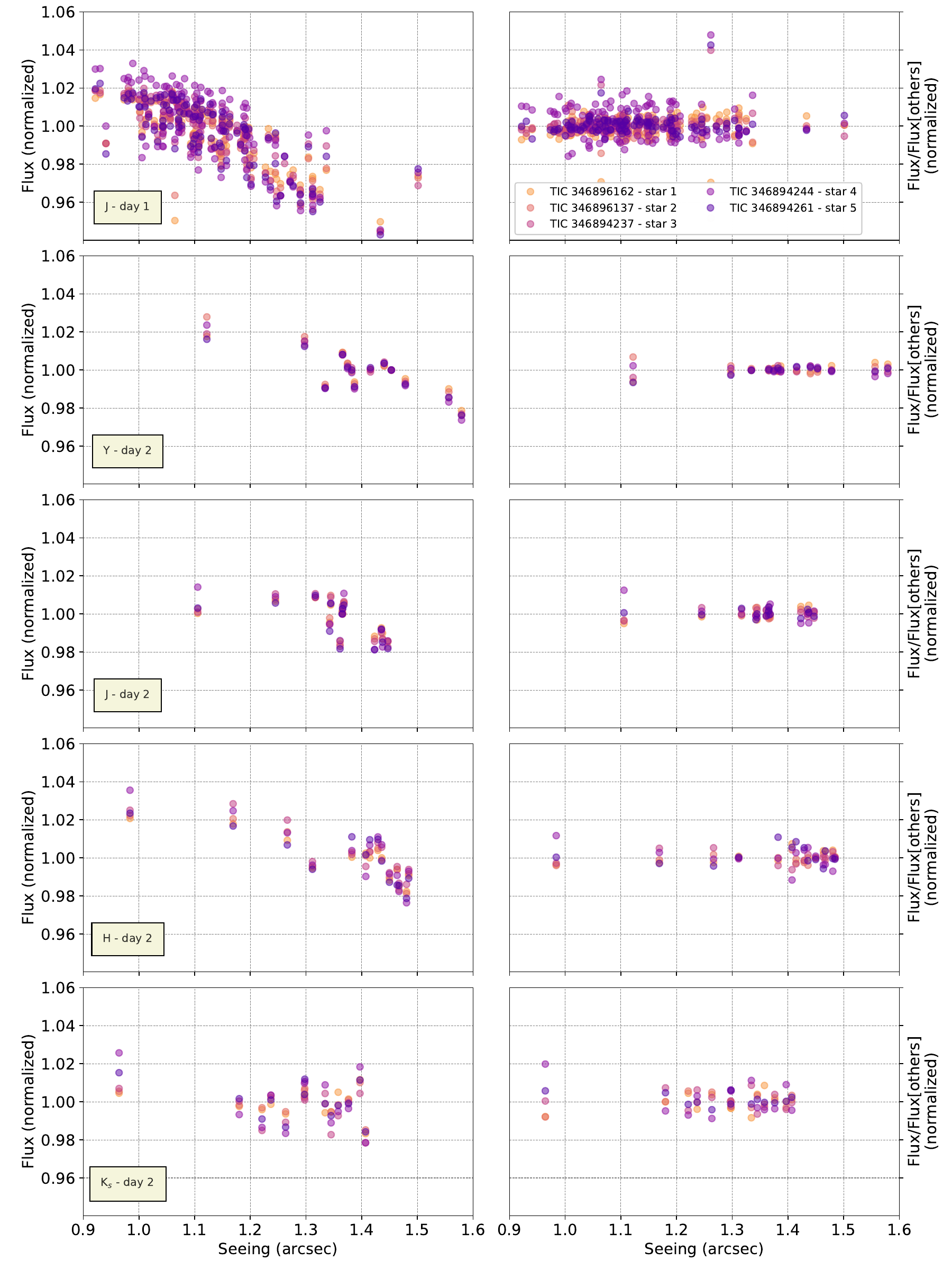}
\caption{Correlation of normalized raw fluxes of reference stars against seeing (left). Each reference star flux has been normalized to its median; one sees a $\sim6$\,\% loss in flux between measurements with a seeing of $\sim$0.9$^{\prime\prime}$ and$\sim$1.6$^{\prime\prime}$. To assess if differential biases were present in our photometric measurements, we compared each of the reference stars to the average of all stars (right) against seeing. The calibrated fluxes show no dependency against seeing, which confirms that our reference stars will serve their purpose in calibrating SIMP0136 timeseries.   \label{fig:flux_vs_seeing}}.
\end{figure}

\begin{figure}
\centering
\includegraphics[width=0.8\textwidth]{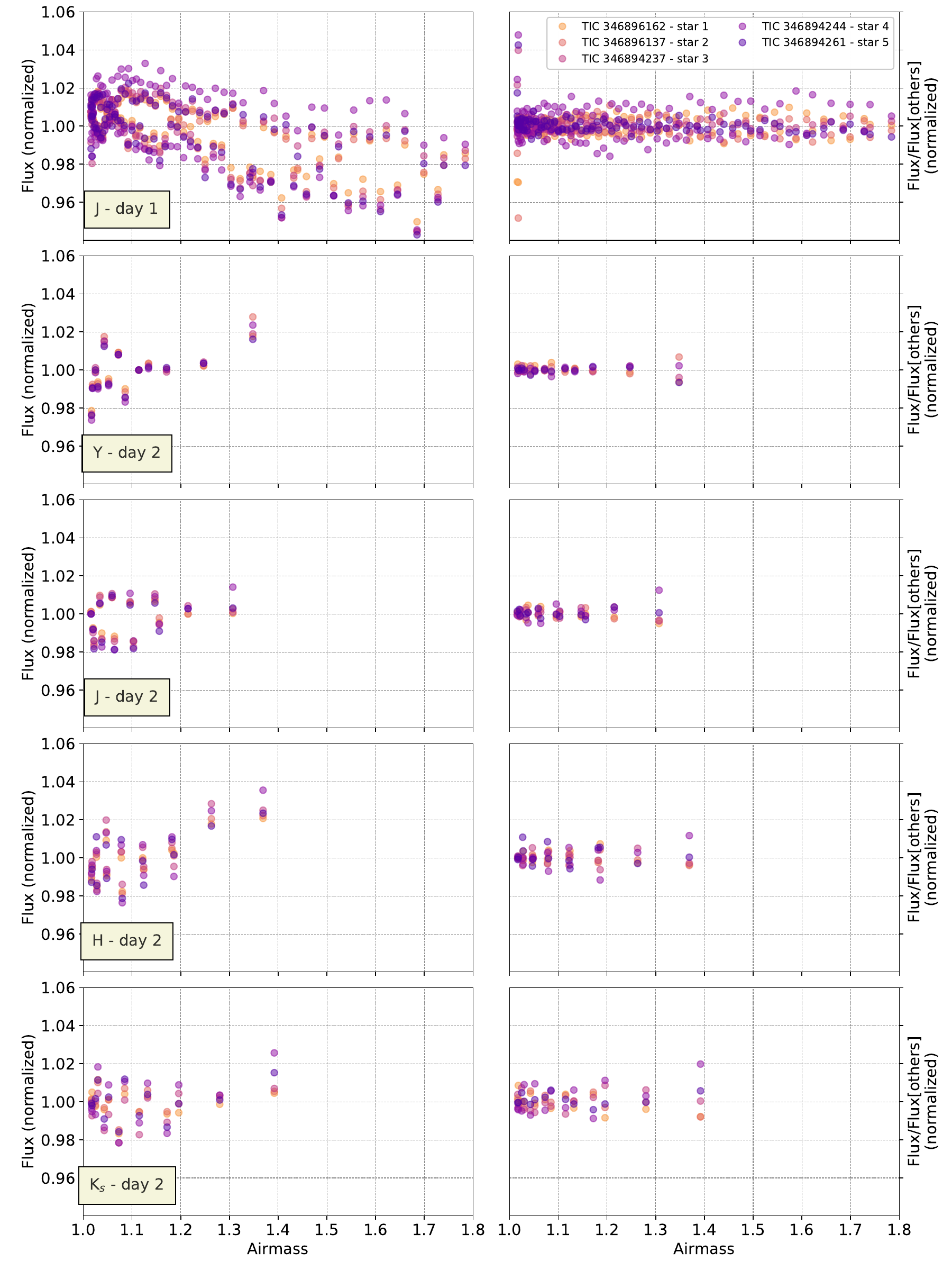}
\caption{Same as Figure~\ref{fig:flux_vs_seeing} but for the dependency of fluxes (raw and corrected) with airmass. \label{fig:flux_vs_airmass}}.
\end{figure}

{We investigated possible correlations between observing conditions and reference star fluxes. Figures~\ref{fig:flux_vs_seeing} and \ref{fig:flux_vs_airmass} illustrate the fluxes measured as a function of seeing and airmass respectively. There is an anticorrelation of flux with both quantities in both day~1 ($J$-only) and day~2 (interleaved $YJHK$) datasets.}

{The purpose of reference stars is to account, at least to first order, for these extinction effects; we, therefore, showed each reference star dependency against seeing and airmass once corrected by the sum of all other reference stars (right panels of Figures~\ref{fig:flux_vs_seeing} and \ref{fig:flux_vs_airmass}). No trend is seen after correction by reference stars at this step and we conclude that reference stars are, for all practical purposes, stable to well within 1\%.}

\begin{figure}
\centering
\includegraphics[width=0.8\textwidth]{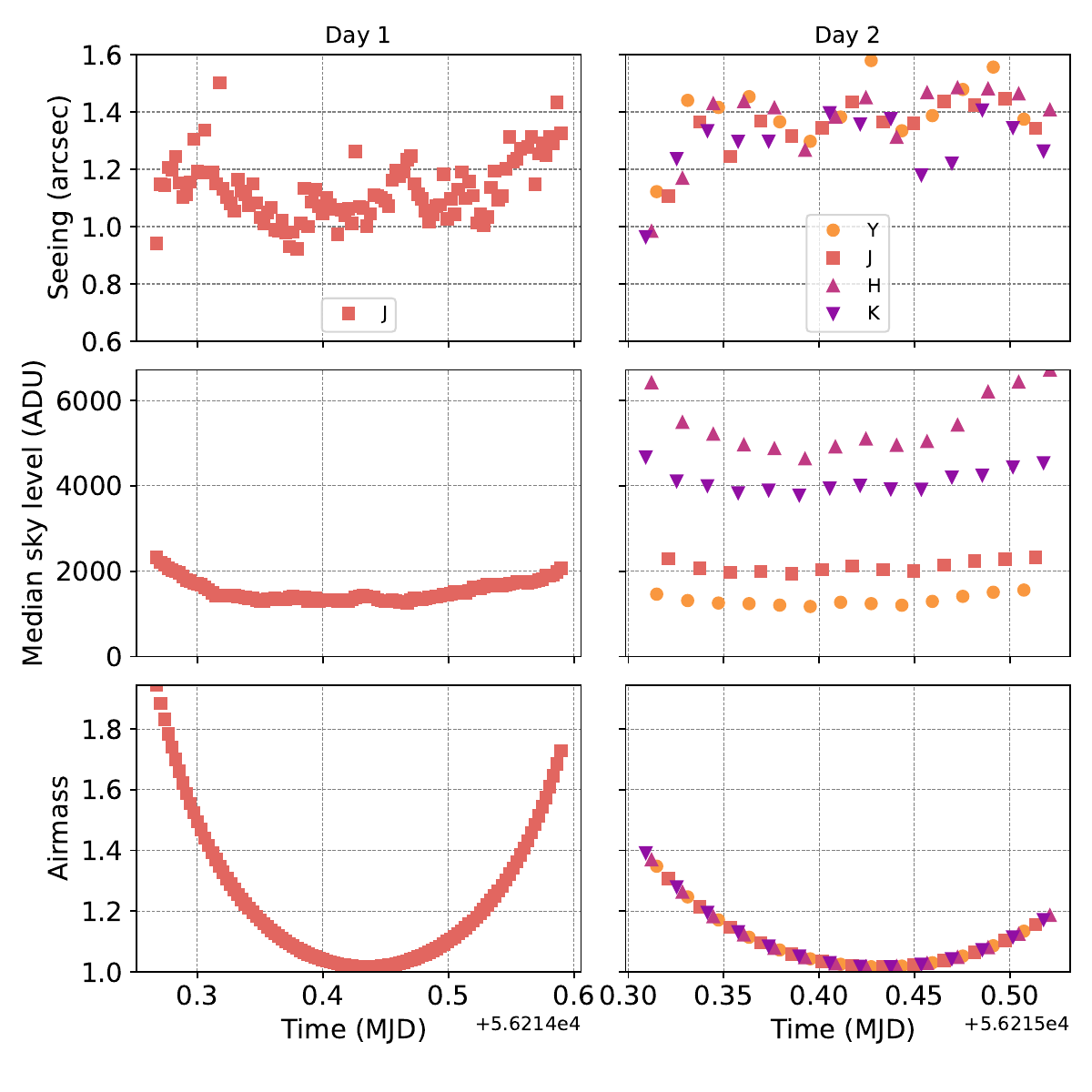}
\caption{Variation of the seeing, median sky background and airmass through the $J$-band observing night. Sky background, and to a lesser extent seeing, are correlated with the airmass through the observing sequence.\label{fig:seeing_sky_airmass}}.
\end{figure}

{ The dependency of raw fluxes against airmass can easily be understood as an increase in atmospheric extinction at higher airmass while the dependency with seeing probably arises from a combination of aperture losses as well as the covariance of seeing with airmass (i.e., a worsening seeing correlates with higher extinction as both happen at a higher airmass).  Figure~\ref{fig:seeing_sky_airmass} shows the changes of seeing, median sky level and airmass through our datasets. The expected mild correlation of seeing and sky background is seen mostly on day~1. Day~2 observations are shorter and explore a smaller airmass range. }

\clearpage
\bibliographystyle{aasjournal}
\bibliography{main}
\end{CJK*}
\end{document}